\documentclass{emulateapj}

\usepackage{lscape}
\usepackage{psfig}

\def\arcsec{$\,^{\prime\prime}$~}

\def\erg/cm2sec{ergs~cm$^{-2}$~s$^{-1}$}  
\def\ergcm2{ergs~cm$^{-2}$}

\newcommand{\lsim }{{\lower0.8ex\hbox{$\buildrel <\over\sim$}}}
\newcommand{\gsim }{{\lower0.8ex\hbox{$\buildrel >\over\sim$}}}

\def\Chandra{${\it Chandra}$}

\def\simge{\mathrel{%
   \rlap{\raise 0.511ex \hbox{$>$}}{\lower 0.511ex \hbox{$\sim$}}}}
\def\simle{\mathrel{
   \rlap{\raise 0.511ex \hbox{$<$}}{\lower 0.511ex \hbox{$\sim$}}}}

\newcommand{\Msun}{\ifmmode {M_{\odot}}\else${M_{\odot}}$\fi}
\newcommand{\Lsun}{\ifmmode {L_{\odot}}\else${L_{\odot}}$\fi}
\newcommand{\Rsun}{\ifmmode {R_{\odot}}\else${R_{\odot}}$\fi}

\shorttitle{Faint X-ray Sources in Terzan 5}
\shortauthors{Heinke et al.}

\begin{document}
\title{Faint X-ray Sources in the Globular Cluster Terzan 5}   

\author{C. O. Heinke\altaffilmark{1,2}, R. Wijnands\altaffilmark{3}, 
H.~N. Cohn\altaffilmark{4}, P.~M. Lugger\altaffilmark{4},  
J.~E. Grindlay\altaffilmark{5}, D. Pooley\altaffilmark{6,7}, 
W.~H.~G. Lewin\altaffilmark{8}}
\altaffiltext{1}{Northwestern University, Dept. of Physics \& Astronomy, 
2145 Sheridan Road, Evanston IL 60208; cheinke@northwestern.edu}

\altaffiltext{2}{Lindheimer Postdoctoral Fellow}

\altaffiltext{3}{Astronomical Institute 'Anton Pannekoek', University of Amsterdam, Kruislaand 403, 1098 SJ Amsterdam, The Netherlands}
\altaffiltext{4}{Dept. of Astronomy, Indiana University, Swain West 319, Bloomington, IN 47405}
\altaffiltext{5}{Harvard-Smithsonian Center for Astrophysics, 60 Garden Street, Cambridge, MA 02138}
\altaffiltext{6}{Dept. of Astronomy, University of California, Berkeley, CA, 94720-3411}
\altaffiltext{7}{\Chandra\ Fellow}
\altaffiltext{8}{Kavli Institute for Astrophysics and Space Research, Massachusetts Institute of Technology, Cambridge, MA 02139}


\begin{abstract}

We report our analysis of a \Chandra\ X-ray observation of the rich globular
cluster Terzan 5, in which we detect 50 sources to a limiting 1.0-6
keV X-ray luminosity of $3\times10^{31}$ ergs s$^{-1}$ within the half-mass
radius of the cluster.  Thirty-three of these have $L_X>10^{32}$ ergs s$^{-1}$, 
the largest number yet seen in any globular cluster.  In addition to the 
quiescent low-mass X-ray binary (LMXB, identified by Wijnands et al.), 
another 12 relatively soft sources may be quiescent LMXBs.  We compare 
the X-ray colors of the harder sources in Terzan 5 to the Galactic 
Center sources studied by Muno and collaborators, and find the Galactic Center 
sources to have harder X-ray colors, indicating a possible difference in the populations. 
We cannot clearly identify a metallicity dependence in the production of 
low-luminosity X-ray 
binaries in Galactic globular clusters, but a metallicity dependence of the 
form suggested by Jord\'{a}n et al. for extragalactic LMXBs is consistent 
with our data.

\end{abstract}


\maketitle

\section{Introduction}\label{s:intro}

Globular clusters are highly efficient at producing X-ray binaries
through dynamical interactions \citep{Ivanova04}. For luminous
low-mass X-ray binaries (LMXBs; for the purposes of this paper, all 
discussion of LMXBs refers to those containing accreting neutron stars),
this has been known for many years, as their production rate per unit
mass is $>100$ times that of the rest of the Galaxy \citep{Clark75}.
Only in the past few years has it been possible to study the
populations of faint X-ray sources in the densest globular clusters in depth,
due to the high spatial resolution of the \Chandra\ {\it X-ray
 Observatory} and optical identifications by the {\it Hubble Space
  Telescope} \citep[see][ for a review]{Verbunt04}.  These 
low-luminosity X-ray sources include quiescent LMXBs (qLMXBs), identified by
their previous outbursts or soft blackbody-like X-ray spectra
\citep{intZand01,Rutledge02a}, cataclysmic variables (CVs) 
generally identified by their blue, variable optical counterparts
\citep{Cool95}, chromospherically active main-sequence binaries (ABs)
identified by their main- (or binary-) sequence, variable optical
counterparts \citep{Edmonds03a}, and millisecond pulsars (MSPs) identified by
their spatial coincidence with radio timing positions
\citep{Grindlay01a}.  

These lower-luminosity X-ray sources are also produced through
dynamical interactions, as demonstrated by the correlation between the
number of X-ray sources in a cluster and its ``collision number''
\citep{Pooley03}, a measure of the cluster's stellar interaction rate.
One of the clusters with the highest collision numbers is Terzan 5, a
dense cluster located 8.7 kpc away, near the Galactic center \citep{Cohn02,Heinke03b}.  
This cluster hosts a transiently luminous LMXB, EXO 1745-248, first
detected through X-ray bursts in 1980 \citep{Makishima81} and
irregularly active since then \citep{Wijnands05a}.  Terzan 5 also
hosts at least 30 MSPs, including the fastest known 
\citep{Ransom05, Hessels06}, the largest 
number yet discovered in any globular cluster.  Terzan 5 also has a 
 high metallicity of [Fe/H]=-0.21 \citep{Origlia04}. 

The incidence of bright LMXBs in globular clusters has been clearly 
associated with increasing metallicity, but to date the effects of 
metallicity on faint X-ray binaries in globular clusters have not been 
studied.  \citet{Grindlay87} identified an apparent trend for LMXBs to be 
more common in metal-rich globular clusters in the Milky Way, 
confirmed for the Milky 
Way and M31 by \citet{Bellazzini95}.  \citet{Kundu02} demonstrated that 
metal-rich clusters are 3 times more likely than metal-poor clusters to 
possess LMXBs in the elliptical NGC 4472.  This result has been confirmed for 
various early-type galaxies by \citet{Maccarone03, Kundu03, Sarazin03} and 
\citet{Jordan04}, the last offering a scaling for the likelihood of a 
cluster in M87 hosting an LMXB of $(Z/Z_{\odot})^{0.33\pm0.1}$.  Suggested 
explanations for this effect are a dependence of the cluster initial mass 
function on the metallicity \citep{Grindlay87}, a change in the rate of tidal 
captures \citep{Bellazzini95}, a change in the strength of stellar winds 
\citep{Maccarone04a}, and a change in magnetic braking rates due to 
differences in convective zone depths \citep{Ivanova06}. 

Terzan 5 has been observed by \Chandra\ in 2000 (two closely spaced observations) and 2002.  
The 2000 observations caught  
EXO 1745-248 during a bright outburst \citep{Heinke03b}.
The high resolution of \Chandra\ allowed the detection of nine 
additional low-luminosity sources within the cluster, and some
useful spectral information of the transient was recovered 
from the readout streak.  In the
2002 observation, EXO 1745-248 was observed at a typical X-ray
luminosity for a qLMXB ($L_X=2\times10^{33}$ ergs s$^{-1}$), but with
an unusually hard X-ray spectrum \citep{Wijnands05a}.
\citet{Wijnands05a} also detected a large number of faint X-ray
sources in Terzan 5, which are the focus of this paper.  


\section{Observations and Reduction}\label{s:obs}
We observed Terzan 5 with \Chandra\ for 39.3 kiloseconds on July 13-14,
2003, using the ACIS-S3 chip (turning off other chips to avoid the
possibility of telemetry saturation in case of an LMXB outburst).   We
reduced and analyzed the data using the  
\Chandra\ Interactive Analysis of Observations (CIAO) v.\ 3.2.1 
software\footnote{http://asc.harvard.edu/ciao/.}.  We reprocessed the
level 1 
event files using the latest (time-dependent) gain files, using bad
pixel files generated with the new ACIS\_RUN\_HOTPIX routine, and
without the pixel 
randomization which is applied in standard data processing.  We 
filtered on grade, status, and good time intervals supplied by
standard processing.  
The later part of the observation suffers from elevated background
levels.  We removed 4.0 kiloseconds (ksec) of data with background flares, 
for a total good time of 35.3 ksec.


\subsection{Detection and source property extraction}\label{s:extr}
We focus our analysis upon the sources (with one exception that we discuss below) within
the cluster half-mass radius (r$_h$=0\farcm83, Harris 1996, updated 2003),
as done for other clusters \citep[e.g. ][]{Pooley03, Heinke05a}. 
This offers an excellent balance between 
including most cluster sources and excluding most background sources. Since
globular cluster X-ray    
sources are generally more massive than the typical cluster star, they
tend to concentrate towards the core of dynamically relaxed
clusters such as Terzan 5.  

\begin{figure}
\figurenum{1}
\epsscale{1.3}
\plotone{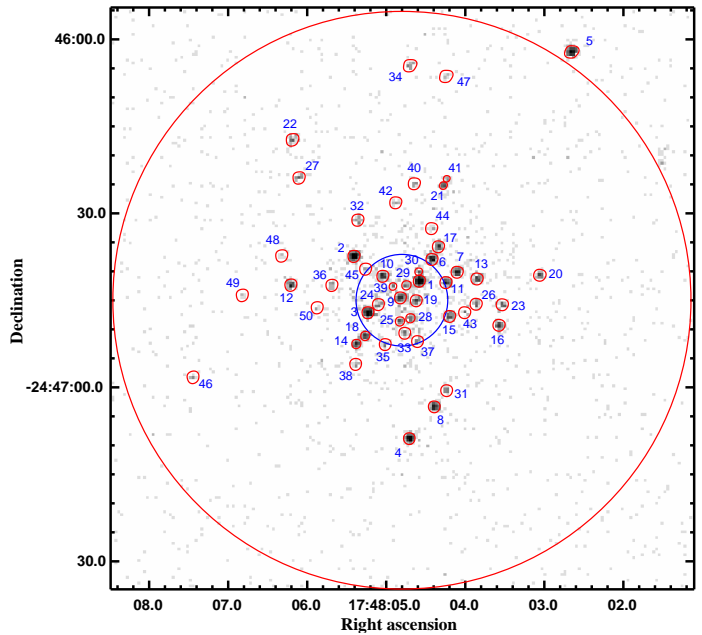}
\caption{ X-ray image of Terzan 5, with the core and half-mass radii 
  marked, and the extraction regions for the sources associated with
  Terzan 5.  
} \label{fig:halfbox}
\end{figure}

  We selected an energy band of 0.5-7.0 keV to 
search for sources with maximum sensitivity while minimizing the 
background.  We ran two wavelet detection algorithms, the CIAO task
WAVDETECT \citep{Freeman02}, and the 
PWDETECT\footnote{http://www.astropa.unipa.it/progetti\_ricerca/PWDetect/} 
algorithm 
\citep{Damiani97}, on ACIS chip S3, with broadly similar results.  We
find that PWDETECT is somewhat more effective at 
identifying faint sources very near to brighter sources, while
WAVDETECT is more reliable over large fields; and thus adopt PWDETECT
results within r$_h$, and WAVDETECT results otherwise.
We choose the sensitivity of our   
detection algorithms to identify no more than one spurious 
source within 1 r$_h$ (for PWDETECT) and the S3 chip (for WAVDETECT).  We
find a total of 49 sources at or within the cluster r$_h$.
One bright source (CXOGLB J174802.6-244602) located just beyond 1 r$_h$ 
seems highly likely to be associated with the cluster (due to its high flux; its X-ray colors are consistent with either an AGN or CV), and we include it also
in our analyses of the cluster sources.  We tabulate properties of the likely
cluster sources in Table 1, and other sources in the field in
Table 2.  In Figure \ref{fig:halfbox}, we show a 0.5-7 keV \Chandra\ 
image of Terzan 5, including the extraction regions for our identified sources. 

To extract source photometry and spectra and improve source positions,
we use the IDL program ACIS\_EXTRACT \citep{Broos02,
  Feigelson02}. This program automates the extraction process using
CIAO and FTOOLS software, and is designed for working with crowded
fields and multiple observations.  The principal benefit of this
software for us was its production of extraction regions designed to
match the contours for a point-spread function (PSF) fraction of the
user's choice, and 
application of this PSF fraction to results from photometry and
spectroscopy.  We briefly describe the major elements of this
process. 

We refined the positions (starting from the PWDETECT positions) of 
all sources within the cluster by finding the centroid of the 
detected counts within the ACIS\_EXTRACT-produced extraction radii.
  We extracted counts from within the 90\% contour
for most sources, choosing the 95\% contour for relatively bright and
isolated sources, and smaller contours for faint sources experiencing
heavy crowding.  We note that the sources within the cluster core radius 
are likely to suffer some degree of confusion.   We compute
background-subtracted photometry for each source, accounting for the
fraction of the PSF enclosed, and for the energy dependence of this
fraction.  To compute fluxes (for the 0.5-2.5 and 1.0-6 keV bands), we
find the countrates observed 
in several narrow bands (0.5-1, 1.0-1.5, 1.5-2.0, 2.0-2.5, 2.5-3.3,
3.3-4.5, 4.5-4.7, 4.7-6.0 keV) and compute conversion factors from
photon fluxes to unabsorbed energy fluxes for each band \citep[assuming
$N_H=1.2\times10^{22}$ cm$^{-2}$,][]{Cohn02}, then sum the energy fluxes.  
We extract
background spectra using regions sized to include $>100$ counts and
excluding mask regions around identified sources.  Response matrices
are constructed using the CIAO tool MKACISRMF, and effective area
files are modified to account for the energy-dependent aperture
corrections \citep[by ACIS\_EXTRACT, see ][]{Broos02}.  

The substantial column density in the direction of Terzan 5
($1.2\times10^{22}$ cm$^{-2}$) is likely to obscure
background AGN and reduce their number counts.  We expect 
approximately 8 background AGN with more than 10 counts on the S3 chip
 outside two r$_h$ from Terzan 5 \citep[based on the results of ][and including the effects of absorption]{Giacconi01}, but we see 22 such
 sources.  This indicates a population of Galactic sources, including
 foreground stars and possible faint CVs in the galactic bulge
 \citep{Grindlay02}.   Therefore we estimate our noncluster source
 numbers by looking at the radial distribution of X-ray sources.
 Outside a radius of two r$_h$ from Terzan 5, the numbers of X-ray
 sources above 10 counts are at 0.4 arcminute$^{-2}$, implying that
 $\sim$one of 37 such sources within 1 r$_h$ is not
 associated with the cluster.  For the 1-2 r$_h$ annulus, 3.4 of 6
  sources above 10 counts are likely to be associated with the cluster, and
 for the 2-3 r$_h$ annulus only 1.6 of 6 sources may be cluster
 members.  Foreground chromospherically active stars may be
 identifiable through optical counterparts or very soft spectra, while
 background AGN and distant CVs may be indistinguishable.

\section{Analysis}\label{s:analysis}

\subsection{Astrometry and possible counterparts}

\begin{figure}
\figurenum{2}
\epsscale{1.1}
\plotone{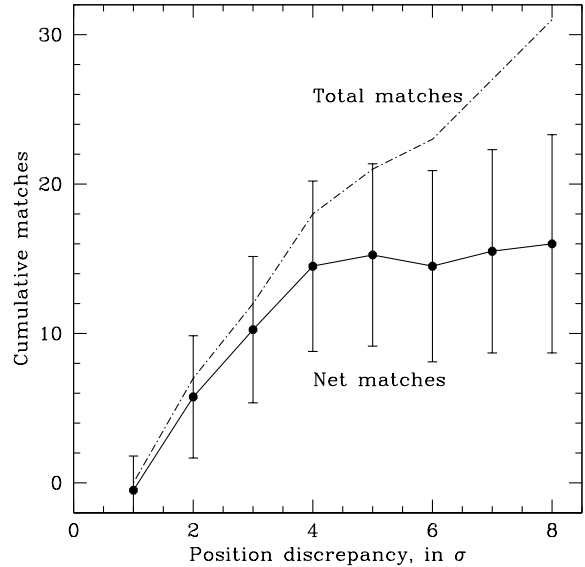}
\caption{ Numbers of possible optical counterparts outside Terzan 5 
half-mass radius (total, dash-dotted line; and net, after subtraction 
of average number of matches with 15'' shifts, solid) vs. search radius 
(in units of $\sigma$, combined \Chandra and USNO errors).  
} \label{fig:matches}
\end{figure}

We searched for possible optical counterparts to sources 
outside r$_h$ using the USNO B1.0 catalog \citep{Monet03}.  We
identify several probable optical counterparts, of which we take 
six relatively secure and uncrowded matches (marked with * in Table 2)
 to define our reference
frame.  To match the USNO B1.0 frame, we shift our \Chandra\ positions
by +0\farcs083 ($\pm$0.003) in right ascension and -0\farcs175 ($\pm$0.004)
 in declination.  Defining $\sigma$ as the sum in quadrature of the USNO 
uncertainties and the centroiding errors derived from ACIS\_EXTRACT, we 
measured the numbers of possible counterparts within 1$\sigma$ to 8$\sigma$, 
and compared these to the numbers of spurious counterparts found by shifting 
the \Chandra\ positions 15\arcsec in four directions 
(Fig. \ref{fig:matches}). 
(We exclude faint USNO stars with large quoted errors of 0.999\arcsec.)  
We find that real 
matches occur for separations up to $\sim4\sigma$, within which 
14.5$\pm$5.7 of the 18 total matches represent an excess over the expected number of spurious matches.  
Systematic errors (due to optical crowding),
unmeasured proper motions, and binaries with fainter stellar companions
may contribute to generating these apparently large errors. 

\begin{figure}
\figurenum{3}
\epsscale{1.25}
\plotone{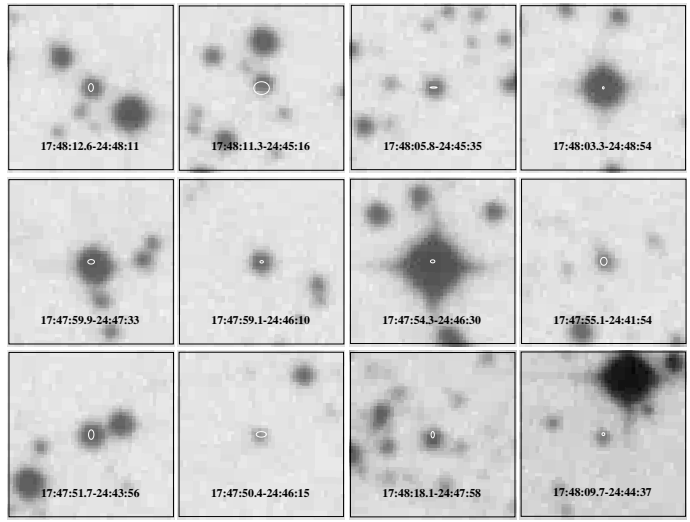}
\caption{ Likely optical counterparts outside Terzan 5 half-mass radius.  
Ellipses represent twice the summed USNO and \Chandra\ uncertainties.  Each 
box is 33\arcsec on each side.  Optical images are from UKST $B$ image, 
except CXOU J174818.1-244758 for which the UKST $I$ image is used.
} \label{fig:12match}
\end{figure}

 We carefully
inspected UKST $B$ and $I$ plates of the region\citep{Hambly01}\footnote{SuperCOSMOS via Aladin, http://aladin.u-strasbg.fr/aladin.gml}.  
We identify 12 very likely optical counterparts, for which we show postage 
stamps in Fig. \ref{fig:12match}, and three other possible matches.  

 These 17 matches are marked as
``c'' in notes to Table 
2, including two counterparts discussed in \citep{Heinke03b}.  
Three other possible matches are marked as
``?''.  We note that the sources with optical counterparts include
the six brightest sources (beyond Terzan 5's r$_h$) below 2 keV, and
that their spectra are generally much softer than those of other
sources, as expected for typical stars.

\subsection{X-ray Color Magnitude Diagram}

\begin{figure}
\figurenum{4}
\epsscale{1.25}
\plotone{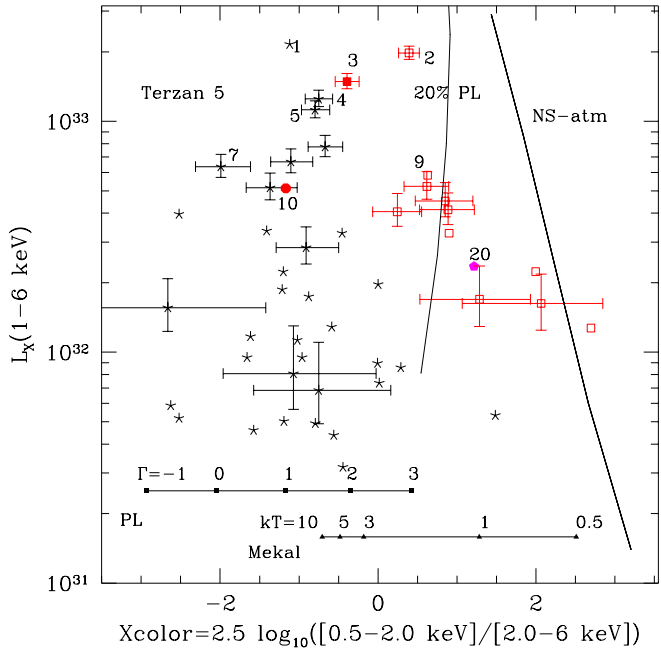}
\caption{ X-ray CMD of Terzan 5, plotting X-ray luminosity (1.0-6 keV,
  corrected for extinction of $1.2\times10^{22}$ cm$^{-2}$) against
  an X-ray color (2.5 log([0.5-2.0 keV]/[2.0-6.0 keV]), thus hardening to the 
left) for sources within half-mass radius of Terzan 5 
(plus CXOGLB J174802.6-244602, see text).  
X-ray luminosities are computed assuming a 6 keV MEKAL
  spectrum to convert photon fluxes into energy fluxes for each of 8
  narrow bands, and summed.  Several sources of particular interest
  (including the X-ray sources identified by \citet{Heinke03b}) are
  identified with their number.  The filled square represents a probable qLMXB, 
the open squares possible qLMXBs, the filled circle a likely millisecond pulsar,
 the magenta pentagon a possible foreground star or qLMXB (see \S 3.5.3), 
and five-pointed stars sources of unknown nature.  See text for 
description of model tracks.
} \label{fig:XCMD}
\end{figure}

X-ray color-magnitude diagrams (XCMDs), plotting an X-ray color against
detected counts or X-ray luminosity, have been used by several authors as an
effective way to understand the source populations in globular
clusters \citep{Grindlay01a,Pooley02a,Heinke03d}.  
Due to the high absorption towards Terzan 5, several X-ray sources
have no detected counts below 1.5 keV.  To avoid displaying large
numbers of upper limits, we therefore choose a nonstandard X-ray
color, 2.5 log([0.5-2.0 keV]/[2.0-6.0 keV]). The extrapolation from a few 
detected 0.5-1 keV counts to the 0.5-1 keV luminosity is extremely uncertain, 
so we choose to compute and plot more accurate 1-6 keV luminosities.  
This reduces the size of the luminosity errors by up to 90\% in some cases.
  We plot the XCMD in Figure 
\ref{fig:XCMD} vs. the 1.0-6.0 keV X-ray luminosity inferred (assuming
$N_H=1.2\times10^{22}$ cm$^{-2}$) for each source (see above).  

To make our XCMD useful for comparison with other clusters, we plot on
this XCMD the X-ray colors of power-law (PL) and thermal plasma \citep[MEKAL in
XSPEC, ][]{Liedahl95} spectral models.  We also plot the expected
location of neutron stars with hydrogen atmospheres radiating away
stored heat \citep[labeled NS-atm; ][]{Rybicki05}, as expected for
quiescent LMXBs (qLMXBs) containing neutron stars (NSs).  Quiescent
LMXBs are often observed to show a second, harder spectral component generally
fit with a PL of photon index 1-2.  This component may make up
anywhere from $<5$\% to the majority of the detected X-ray flux
\citep{Rutledge01a,Heinke03a,Wijnands05a}.  We plot the effect of
including such a harder PL component (with photon index 1.5), that 
makes up 20\% of the 0.5-6 keV flux, in Fig. \ref{fig:XCMD}.

We note immediately that there are very few X-ray sources lying along
the track for NS atmospheres in Fig. \ref{fig:XCMD}.  However, there are
substantial numbers of sources between this track, and the track for NS
atmospheres plus a 20\% PL component.  Objects with this range of
X-ray colors at the extinction of Terzan 5 would require PL indices
greater than 3, or MEKAL temperatures less than 2 keV.  Among the
globular clusters for which excellent optical and ACIS data are available 
(47 Tuc, \citet{Grindlay01a}, \citet{Edmonds03a}; 
NGC 6397, \citet{Grindlay01b}; $\omega$ Cen, \citet{Cool02b}, \citet{Gendre03a}; 
NGC 6752, \citet{Pooley02a}; M4, \citet{Bassa04}; NGC 288, \citet{Kong06}; M30, \citet{Lugger06}), 
there are no examples of CVs of
similar brightness and very soft spectrum\footnote{X10 in 47 Tuc is the softest bright CV we know of, but still has a powerlaw photon index less than 3.}, leading us to suspect that these
sources are probably qLMXBs.  
Several authors have suggested that most globular cluster qLMXBs appear softer 
than CVs, even allowing for a substantial powerlaw contribution. 
\citet{Pooley06a} plot a modified XCMD for all globular cluster sources 
(their Fig. 1), which shows that nearly all identified globular cluster 
CVs have spectra harder than CX3.
We note that some bright qLMXBs 
have relatively hard spectra \citep[see ][and below]{Wijnands05a}, and
that there may be large populations of faint qLMXBs with hard spectra
\citep{Jonker04,Heinke05b}.  However, we think it likely, based on
studies of other clusters, that the
X-ray sources we have labeled with open boxes 
in Figure \ref{fig:XCMD} are mostly
qLMXBs, and that they represent the majority of the qLMXBs in that luminosity
range.  The high extinction towards Terzan 5 makes spectral analysis
of these sources difficult, and optical studies with current
instrumentation nearly impossible, so it may be a long time before
the X-ray sources in Terzan 5 can be conclusively identified.  

\subsection{X-ray Color-Color Plot}

We can also study the X-ray colors of the X-ray sources in Terzan 5 by
producing a color-color plot.  Our motivation is to compare the X-ray
colors of Terzan 5 sources, in relatively hard X-ray bands, to the
faint X-ray sources discovered in the Galactic center by
\citet{Muno03}.  It has been suggested that the relatively bright 
($L_X\sim10^{32-33}$ ergs s$^{-1}$) hard X-ray sources
in globular clusters are largely composed of ``intermediate polars''
or DQ Her stars \citep{Grindlay95,Edmonds99}, CVs in which the
accretion flow is 
channeled by the magnetic field of the white dwarf onto its magnetic
poles \citep{Patterson85}.  It has
also been suggested that the bright hard X-ray sources at the Galactic
center are intermediate polars \citep{Muno03}.  We will test whether
the X-ray sources in these two environments show similar X-ray spectra
in a band where interstellar extinction is not very important ($>2$
keV). 

\begin{figure}
\figurenum{5}
\epsscale{1.25}
\plotone{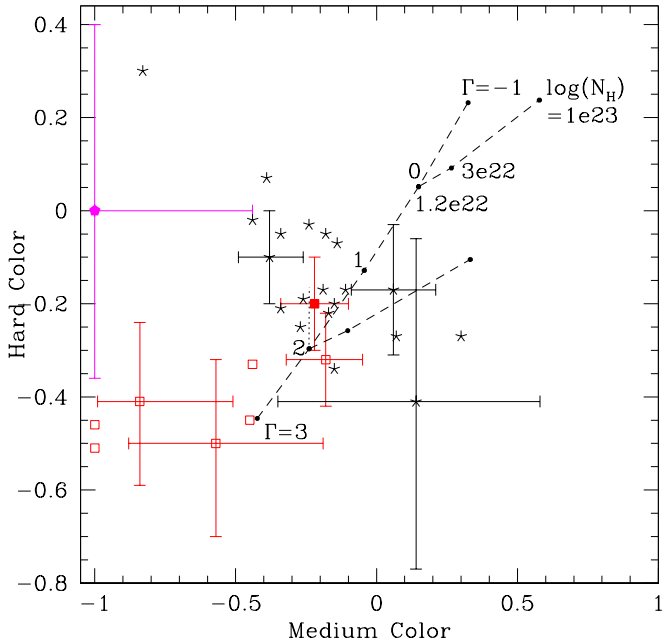}
\caption{ X-ray color-color plot for Terzan 5, showing sources
  associated with the cluster and having more than 20 counts.  The
  dashed tracks indicate colors for a power law of different spectral indices,
  and (shifting to right) for PL models of fixed spectral index
  with increasing $N_H$. The dotted line ascending from the $\Gamma=2$ powerlaw
  model indicates the impact of adding a 6.5 keV Fe line of up to 1.6 keV
  equivalent width.  See text for definitions of colors.
} \label{fig:color}
\end{figure}

We use the same bands and colors as Figure 12 of \citet{Muno03}, with
colors of the 
form $(h-s)/(h+s)$, $h$ and $s$ being the numbers of counts in the
harder and softer bands respectively.  The medium color uses the bands
3.3-4.7 keV and 2.0-3.3 keV, while the hard color uses the bands
4.7-8.0 keV and 3.3-4.7 keV.  In Figure \ref{fig:color} we plot the
locations of 
Terzan 5 X-ray sources with more than 20 counts, and overlay the
colors of several spectral models.  Since the detectors (ACIS-I
vs. ACIS-S) and absorption
columns are different, the colors are not directly comparable,
thus it is important to plot colors for various models.  Comparison of Figure 
\ref{fig:color} with Figure 12 of \citet{Muno03} shows that a
photon index of 1-2 is a reasonable description of the
majority of 
the Terzan 5 X-ray sources, while the Galactic Center sources are
better described by a median photon index near 0.  A portion of the Galactic
Center sources are even harder.  This indicates that the Galactic 
Center sources may be substantially different from the hard sources in 
Terzan 5. (This question will be investigated in more detail in a 
forthcoming paper.)  We note that one very hard 
bright source (photon index=0.2) has been identified in the globular cluster 
Terzan 1 \citep{Cackett06}.

\subsection{Spectral Fitting and Time Variability}

We extracted spectra and associated files for the brighter cluster
sources as described in \S \ref{s:extr} above.  
We performed spectral fitting for sources with more than 60 counts.
To allow fitting with faint sources, we perform binning using 10 counts/bin for
sources with more than 70 counts, and 8 counts/bin for four sources with
$\sim$60-70 counts.  We also performed spectral fits with 20 or 5 counts/bin,
 and using the C statistic instead of the $\chi^2$
statistic. We found that the results from each method were comparable, with 
our preferred binning giving slightly tighter constraints in several cases.

We choose three models designed to cover the range of spectra
 typically seen in globular cluster X-ray sources in this luminosity range, all
 absorbed by $N_H\geq1.2\times10^{22}$ cm$^{-2}$ and a dust column
 appropriate for $A_V=6.7$.  For the latter we use the {\it scatter}
 XSPEC code kindly provided by P. Predehl \citep{Predehl03}.   Our
 continuum models 
 include a thermal plasma MEKAL spectrum (which may be physically
 appropriate for CVs or ABs), a simple power law (physically
 appropriate for synchrotron radiation from bright MSPs), and a
 two-component model consisting of a hydrogen-atmosphere NS model
 \citep{Rybicki05} plus a
 powerlaw (physically appropriate for qLMXBs).  To permit interesting  
constraints on some parameters for this model, we fix the NS radius (10 km),
 mass (1.4 \Msun), and distance (8.7 kpc), and also fix the slope of
 the associated PL model to $\Gamma=1.5$, a typical slope for
 hard powerlaw components in qLMXB spectra \citep{Rutledge01a}.  

We find, in contrast to other globular clusters, that none of the
bright sources are spectrally consistent with a simple
hydrogen-atmosphere model.  Some relatively soft sources, with
effective PL photon indices greater than 2, can be modeled as
the combination of a hydrogen-atmosphere model and a harder PL
component, as often seen in galactic qLMXBs
\citep[e.g.][]{Campana98a,Rutledge01a}. This is counter to the
suggestion by \citet{Heinke03d} that globular cluster qLMXBs do not
possess this harder component unless they have recently been active, as 
there is no evidence for outbursts by more than one LMXB in 
Terzan 5 in the past 30 years 
\citep[but see ][ for a discussion]{Wijnands05a}.  

We performed a Kolmogorov-Smirnov test for variability 
(implemented in ACIS\_EXTRACT)
on the event files for each source associated with Terzan 5.  Those
nine sources which indicate variability with $>95$\% confidence are
indicated in Table 1 with V?; the three sources which have $>99$\%
confidence are indicated with V. Since we are testing 50 sources, we
may expect that two sources might be spuriously identified as variable
for a confidence limit of 95\%; on the other hand, faint sources
are unlikely to be identified as variable, so the above estimate is
overly conservative.  

 Some sources can be identified as
variable between the 2000 and 2002 observations of Terzan 5, and are
indicated with a Y in the table.  
For those sources which were detected in both 2000 and 2003, we have
extracted spectra from the 2000 observation to test whether the
sources require variability.  We
reprocessed the 2000 
data in the same manner as the 2003 data.  We extracted spectra from
1\arcsec circles, and extracted background for most sources from
2\farcs5 annuli around these circles that do not overlap other 
sources (for a few sources we carefully chose alternate background regions). 
Spectra were grouped with 10 or 20 counts per bin.  We note that the
data quality in 2000 is much poorer, due to the high background 
from the transient outburst.
For bright 2003 sources not detected in 2000, we followed the
procedure of \citet{Heinke05a} \citep[see also][]{Muno03} to identify
variability at the 3$\sigma$ level from nondetections.  

\subsection{Discussions of individual sources}

\subsubsection{EXO 1745-248=CXOGLB J174805.2-244647=CX3}
This source, the quiescent counterpart to the bright transient LMXB,
has already been discussed by \citet{Wijnands05a}.  Our
analysis agrees that the spectrum is dominated by a hard PL component.  In
addition to the models in the table, we fit a NS+PL model in which the
PL index was allowed to vary.  This model produced a NS
temperature of $91^{+40}_{-91}$ eV, giving the NS component
$13^{+33}_{-13}$\% of the total unabsorbed 0.5-10 keV flux.  

CX3 is not identified as clearly variable during the 2003 observation,
with a K-S probability of nonvariance of 6.6\%.  This finding differs
from that of \citet{Wijnands05a} (less than 5\% probability of
nonvariance), but this small difference is likely due to
different (but equally justifiable) choices of data reduction procedures.

\subsubsection{CX2=CXOGLB J174805.4-244637=W3}
\begin{figure}
\figurenum{6}
\epsscale{1.2}
\plotone{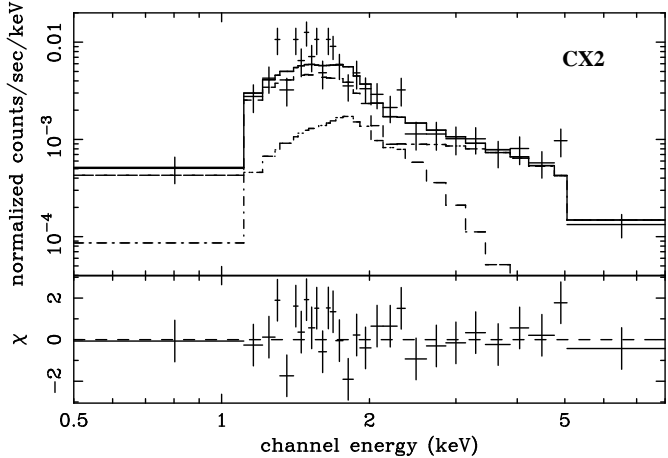}
\caption{ X-ray spectrum of CX2, showing the data (crosses), model 
(solid line), and the portion of the model due to each component 
(hydrogen atmosphere model as the dashed line, power-law component 
as the dot-dashed line). 
} \label{fig:cx2}
\end{figure}

This source, observed by \citet{Heinke03b} and labeled W3, is the
second brightest source in Terzan 5.  Its spectrum (Figure \ref{fig:cx2})
is the clearest example of a two-component spectrum among our
sources.  Adding a NS atmosphere component to an absorbed PL fit
gives an F-test probability of 3.3\% that the improvement in statistic
could have arisen by chance.  We regard this source as an almost certain
qLMXB. 

It is clear from the recorded counts and inferred X-ray luminosities
in Table 1 compared to \citet{Heinke03b} that CX2 has varied in
brightness between 2000 and 2003.  We confirm this by simultaneously
fitting spectra of CX2 from 2000 and 2003. CX2's 2000 flux was at
26$^{+7}_{-7}$\% of its 2003 flux, determined from scaling a PL 
fit.  For our preferred NS
atmosphere plus PL model, acceptable fits (null hypothesis
probabilities, or nhp, $>10$\%) can be obtained by
 varying the NS atmosphere temperature, varying the PL 
normalization, or varying the $N_H$ column, while keeping the other
parameters fixed.  CX2 did not demonstrate clear variability during
the 2003 observation.

\subsubsection{CX20=CXOGLB J174803.0-244640}
This source has an unusually soft spectrum, with substantial flux
below 1 keV.  Our method of computing X-ray luminosity, based on an
assumption of $N_H=1.2\times10^{22}$ cm$^{-2}$, assigns a very high
0.5-1.0 (and thus 0.5-2.5) keV luminosity to this object due to 
these soft photons.  However, our 
spectral analysis of this source reveals that no continuum spectral model can
reproduce the spectral shape if the cluster $N_H$ is assumed.
We find $N_H$ best fit at $6\times10^{20}$ cm$^{-2}$ for
either MEKAL or PL models, with the 90\% confidence upper limit
being 3 or $5\times10^{21}$ cm$^{-2}$ respectively.
  This causes us to suggest that CX20 may be a foreground star.  

Several additional pieces of information support our assertion. 
We note that this source lies outside the cluster core (see Figure 1),
as is likely for a randomly placed foreground object.  CX20 displays
strong variability, with 25 of 40 photons being received within 2.3
(of 35.3 total) ksec.  Our K-S test for variability gives a
$5\times10^{-4}$\% probability of such a distribution by
chance.  Such flaring behavior suggests a stellar flare of a coronally
active star.

A possible optical counterpart appears on the SuperCOSMOS UKST blue
($B_j$) plates available through the
Aladin\footnote{http://aladin.u-strasbg.fr/aladin.gml} image server. 
The USNO B1.0 catalog \citep{Monet03} identifies a nonstellar object
(presumably due to crowding) at 17:48:03.04,-24:46:40.9 (quoted
position errors 0\farcs5), with R=12.57,
B=18.62. This position is consistent with CX20 within the USNO errors. 
However, the color of the star is too red to be consistent with a foreground 
star at the $N_H$ measured above, so unless the USNO magnitudes are 
in error due to crowding, this star is unlikely to be the stellar 
counterpart.   

\subsubsection{Other potential qLMXBs}
The following sources have rather soft X-ray colors, indicating effective 
PL photon indices larger than 3.  A PL 
photon index larger than 3 is rarely observed among non-qLMXBs in this
$L_X$ range; a likely physical explanation is the presence of a blackbody-like NS atmosphere component. 

{\it CX9=CXOGLB J174804.8-244644=W4:} This source exhibits a soft
spectrum, with an inferred PL photon index of
3.62$^{+1.88}_{-1.09}$.  The steep spectrum suggests that it is
dominated by a NS surface, but an absorbed NS atmosphere model alone
produces a relatively poor fit (nhp=7\%) with clear residuals above 3
keV.  Adding a NS atmosphere component to a PL fit allows the
powerlaw to be less steep (best fit $\Gamma=2.2$), but an F-test does
not indicate the NS component is required to improve the fit.  We
designate this source as a possible qLMXB.  Simultaneous fits to the
2000 and 2003 spectra do not require variability. 

{\it CX12=CXOGLB J174806.2-244642=W2:} This source shows an inferred
PL photon index of 3.26$^{+1.27}_{-0.48}$, also suggesting a NS
surface. Like CX9, a NS atmosphere alone is a poor fit (nhp=1\%), and
adding a NS atmosphere component does not substantially improve the
PL fit, although it allows a less steep power law.  We designate CX12 as
another possible qLMXB.  

Simultaneous fits to the 2000 and 2003
spectra suggest that CX12 may have been fainter in 2000 (best fit
gives 2000 flux at 42\% of 2003 flux), but the errors are large enough
that variability is not required at 90\% confidence.

{\it CX14=CXOGLB J174805.3-244652:} This source shows an inferred
PL photon index of 2.28$^{+0.85}_{-0.72}$, perhaps not as soft as other
suggested qLMXBs.  A NS atmosphere alone is
again a poor fit, but adding a NS component does improve the PL 
fit (an F-test gives an 8.5\% probability that this improvement is due
to chance).  This is a marginal candidate for a qLMXB.  

CX14's location 6\arcsec from EXO 1745-248 during its 
2000 outburst prevents measurement of possible long-term
variability. We identify variability, with 98.2\% confidence, during
the 2003 observation.

{\it CX15=CXOGLB J174804.1-244647=W8:} This source shows an inferred
PL photon index of 3.45$^{+1.47}_{-0.64}$, suggesting a NS
component.  Again, a NS atmosphere alone is a poor fit, but adding a
NS component improves a PL fit (an F-test gives a 6.8\%
probability of such a chance improvement).  CX15 is a good candidate
for a qLMXB.  Simultaneous fits to 2000 and 2003 data reveal no
evidence for variability.

The above sources are reasonable candidates for qLMXBs.  Several other
X-ray sources are softer than these sources, but have fewer than 60
counts, making spectral fitting difficult.  We think these softer
sources reasonable candidates for qLMXBs, probably with smaller
PL spectral components (see Figure \ref{fig:XCMD}). 
The level of certainty in classification of qLMXBs that can be
attained in other clusters has not been reached for these sources,
with the exception of CX3 and CX2.  In addition, the luminosities of
these candidate qLMXBs must be regarded as extremely poorly
determined, since their inferred luminosities depend upon
extrapolation from a few counts below 1.5 keV.  This makes comparison
of the source content of Terzan 5 
with other clusters rather uncertain.

\subsubsection{Other sources seen in 2000}

A number of X-ray sources appear to be fainter in 2000 than in 2003, 
as determined
from fitting an absorbed PL model to both spectra, with only a normalization 
constant allowed to vary between the two spectral fits.    
For CX6 (W5), the normalization of the 2000 data is 33$^{+44}_{-33}$\% that
of the 2003 data. For CX7 (W9), the 2000 normalization is
30$^{+51}_{-30}$\% of the 2003 normalization.  
For CX11 (W7), the 2000 normalization is consistent
(79$^{+79}_{-69}$\%) with that of 2003, also for CX16 (W10,
122$^{+104}_{-91}$\%).  
Only for CX8 (W6) is the best fit normalization of the 2000 data
marginally higher (1.5$^{+0.6}_{-0.5}$) than the 2003 data.  
We find it
rather odd that so many sources were apparently fainter in 2000 than in
2003.  We can rule out possibilities such as incorrect exposure times or 
oversubtraction of background.   
There are roughly twice as many sources with $L_X>10^{32}$ ergs s$^{-1}$ as  
inferred from incompleteness tests on the 2000 data, which we also do not 
fully understand.

\subsubsection{New sources}

Two bright sources are apparent in the 2003 data which would have
been clearly detected at that brightness in 2000.  The brightest
source in our observation, CX1 (CXOGLB J174804.5-244641) has a
3$\sigma$ upper limit of 6.47$\times10^{32}$ in the 2000 observation, 
 a factor of
five lower than its 2003 detection.  CX1 has a rather hard
spectrum, with $\Gamma=1.10^{+0.39}_{-0.26}$, and an unusually high
luminosity of $L_X=3.7\times10^{33}$ ergs s$^{-1}$.  
CX4 (CXOGLB J174804.7-244708) has a 3$\sigma$ upper limit of 2.73$\times10^{32}$
in the 2000 observation, also a factor of five below its 2003
detection. Its spectrum is slightly softer than CX1
($\Gamma=1.59^{+0.41}_{-0.25}$).  Fainter new 2003 sources 
cannot be ruled out in the 2000 data, and readers may persuade themselves
that they see evidence for, e.g., CX10, CX13, and CX19 in the 2000 data. 

\subsubsection{Millisecond pulsars?}

\citet{Ransom05} have identified a large population of radio MSPs in 
Terzan 5, of which some may be detectable X-ray 
sources.  A typical soft-spectrum low-luminosity 
(few $10^{30}$ ergs s$^{-1}$) MSP like those in 47 Tuc \citep{Grindlay02, Bogdanov06} 
would contribute less than 0.5 counts to our dataset.  Positions for 
Terzan 5 A and C \citep{Fruchter00} show no 0.5-7 keV counts, 
allowing 95\% confidence 
upper limits on their flux of 1-3$\times10^{31}$ ergs s$^{-1}$, 
depending on the chosen spectrum.  However, some MSPs are 
brighter with harder spectra, typically those with higher spindown 
luminosities (e.g. PSR 1821-24 in M28, \citet{Becker03}) or those which  
show hard spectra from shocks between the pulsar wind and material from 
the companion (e.g. 47 Tuc-W, \citet{Bogdanov05}, and NGC 6397-A, \citet{Grindlay02}).  Some of these MSPs should be detectable in Terzan 5.  
A preliminary position for one MSP is indeed a very close match to the position
 of the hard X-ray source CX10 (S. Ransom 2005, priv. comm.).   
Other X-ray sources may be identified with MSPs as more positions become 
available. 

\subsection{Spatial distribution of X-ray Sources}

We estimate the ratio of the masses of the X-ray binaries in Terzan 5 
to the masses of stars in Terzan 5, by comparing their radial 
distribution to the radial distribution of cluster stars.  
We use the method described by \citet{Heinke03d} 
\citep[following ][]{Grindlay84} to fit the distribution of X-ray 
sources with a generalized King model, of form 

$S(r) = S_0 [1+(r/r_{c*})^2]^{(1-3q)/2} $

wherein $r_{c*}$ is the optical core radius of the cluster, and $q$ is the 
ratio of the masses of the X-ray sources and the stars that define the 
cluster core radius.  

\begin{figure}
\figurenum{7}
\epsscale{1.2}
\plotone{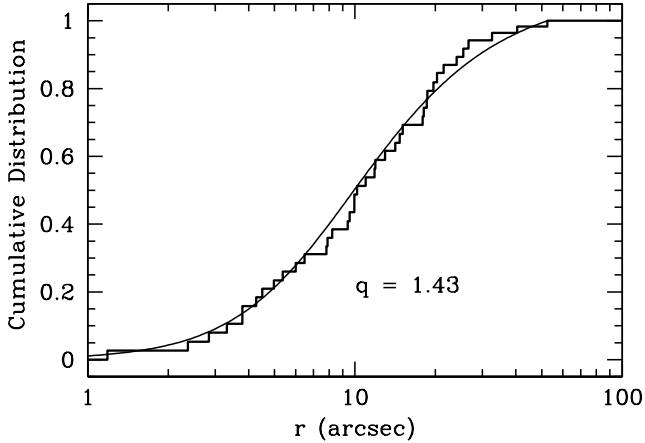}
\caption{ Cumulative radial profile of Terzan 5 X-ray sources with 
$>10$ counts, fitted with our maximum-likelihood generalized King profile. 
} \label{fig:profile}
\end{figure}

We use the distribution of 40 cluster sources above 10 counts (below which we 
are probably substantially incomplete in the core) to measure 
the radial profile of Terzan 5 X-ray sources.  We assume (based on our 
analysis of X-ray sources across the S3 chip, above in \S~\ref{s:extr}) that 
one of these sources is a background (or foreground) source.  
A maximum-likelihood fit to the 
radial profile with our model gives a good fit, with a mass ratio 
$q=1.43\pm0.11$ (Fig.~\ref{fig:profile}).  
This may be compared with the value of $q=1.63\pm0.11$ 
found by \citet{Heinke05a} for 47 Tuc.  It is possible that incompleteness 
in the core may affect our result, but increasing the cutoff value from 10 
counts to 20 ($q=1.45\pm0.14$) or 40 ($q=1.53\pm0.18$) counts does not 
significantly alter the inferred value of $q$.  

For an assumed turnoff mass of 0.9 \Msun \citep[e.g.][]{Bergbusch01}, 
we find a characteristic Terzan 5 X-ray source mass of 1.29$\pm0.10$ \Msun. 
For the subsample of 11 likely qLMXBs plus the known transient LMXB, we 
derive $q=1.64\pm0.25$, not substantially different from the result for the 
entire sample, but consistent with neutron stars and low-mass companions.  
We also note that our lower sensitivity to soft sources (due to extinction) 
may cause us to miss some qLMXBs.  Due to the relatively high mass of NSs 
compared to white dwarfs and cluster stars, this effect may bias our $q$ for
 the total sample downwards, compared to relatively unabsorbed clusters.

\subsection{Luminosity Function and Unresolved Sources}

We choose a limiting luminosity of $L_X$(1-6 keV)$\sim10^{32}$ ergs 
s$^{-1}$ for the following analyses.  For hard sources this is $\sim$15 counts, 
a limit to which we are easily complete.  For soft sources the limit is 
more uncertain.  For the NSATMOS hydrogen atmosphere model alone, a 
5-count detection is $L_X=1.3\times10^{32}$ ergs s$^{-1}$.  However, we have 
found that few if any sources in Terzan 5 are well-described by such a model, 
most requiring a harder power-law component.  Adding a 20\% powerlaw component 
gives 8 counts total, roughly our completeness limit.  We use a limiting 
luminosity of $10^{32}$ ergs s$^{-1}$, but recognize that our results may 
be biased by the loss of soft sources (which is difficult to quantify).

We compute an X-ray luminosity function (XLF) for our sources above this limit 
of the form
$(N > L)\propto L^{-\alpha}$ \citep[following ][]{Johnston96}.  
We find a best-fit slope $\alpha$ of 
0.71$^{+0.25}_{-0.21}$ (errors indicate where the K-S probability falls 
below 10\%); if using the 0.5-2.5 keV luminosities (and a limit of 
$5\times10^{31}$ ergs s$^{-1}$), we find $\alpha=0.44\pm0.03$.  
We note that the 0.5-2.5 keV slope is consistent with the XLF slope found by 
\citet{Pooley02b} for a similar luminosity range in  
 NGC 6440 by the same method, although our 0.5-2.5 keV luminosities are rather
 uncertain.

The X-ray luminosity from unresolved sources can be constrained, once
  the background and the ``spill'' from the PSF wings of identified 
sources are subtracted.  We measured the counts outside the source 
extraction regions within the core and half-mass regions in seven energy bands, 
and subtracted the average background (measured in a large area with no 
bright sources west of the cluster) and the appropriate ``spill'' from 
known sources in each band.  We detect a signal above background in each 
band above 1 keV and below 4.5 keV. We find a total of 96$\pm$19 counts from 
unresolved sources in the core, and 429$\pm$49 counts from unresolved 
sources within the half-mass radius. This translates (using our method 
in \S \ref{s:obs} of computing conversions for each band) to 
$L_X$(1-4.5 keV)$=4.3\pm0.9\times10^{32}$ ergs s$^{-1}$ for the core, and 
$L_X$(1-4.5 keV)$=1.8\pm0.2\times10^{33}$ ergs s$^{-1}$ for unresolved 
sources within the half-mass radius.  If we assume a 7 keV bremsstrahlung 
spectrum, $L_X$(1.0-6 keV)$\sim2.2\times10^{33}$ for the half-mass radius. 
We can constrain the 1-6 keV XLF by comparing the total detected and undetected 
sources with $L_X<10^{32}$ ergs s$^{-1}$ 
(total 1-6 keV $L_X=3.2\times10^{33}$ ergs s$^{-1}$) 
to those with $10^{32}<L_X$(1-6 keV)$<10^{33}$ ergs s$^{-1}$ (total $L_X=8.4\times10^{33}$ ergs s$^{-1}$).  This ratio suggests an XLF index of $\alpha=0.44$ in the 1-6 keV band, somewhat less than inferred above.

\section{Comparisons with Other Clusters}\label{s:comparison}

Terzan 5 has the largest observed number of X-ray binaries above 
$L_X=10^{32}$ ergs s$^{-1}$ of any globular cluster in the
galaxy.  With 28 X-ray sources having inferred unabsorbed  X-ray
luminosities (0.5-2.5 keV) $>10^{32}$ ergs s$^{-1}$ (or 33, for the
1-6 keV range), Terzan 5 contains more than twice as many X-ray binaries
 in this luminosity range as NGC 6440 and 
NGC 6266, the next richest X-ray clusters studied so far
\citep{Pooley02b,Pooley03}.  By comparing results from study of this
cluster to results from other clusters, we can test the dependence of
X-ray source production upon cluster properties, such as the central
density of the cluster.

\subsection{Dependence of encounter frequency on $\rho$, $r_c$, metallicity}  

 \citet{Verbunt87} parametrized the
production rate of X-ray binaries in globular clusters as proportional 
to the square of the central density
$\rho$, and the volume of the core (where most interactions
take place) $\sim r_c^3$, while inversely proportional to the velocity
dispersion in the core.  Thus  $\Gamma\propto \rho^2 r_c^3 /\sigma$.
For a King model globular cluster (and for any virialized cluster), the 
central velocity dispersion should be proportional to $\rho^{0.5} r_c$, 
leading to $\Gamma\propto\rho^{1.5} r_c^2$.  

We use the system of \citet{Heinke03d} \citep[see ][]{Johnston96} to
parametrize the dependence of X-ray binary production on cluster
properties.  This system compares the distribution of X-ray binaries
across a number of clusters with the distribution of production rate
for given dependencies of $\Gamma$ on cluster parameters.  We use
$\Gamma\propto \rho^{\alpha} r_c^{\beta} (Z/Z_{\odot})^{\delta}$, with $\rho$ the central luminosity density, $r_c$ the core radius,\footnote{This is a somewhat simplistic way of treating core-collapsed clusters, which have more complex radial structures and may have a more complex binary history; we defer a more sophisticated analysis.} and $Z/Z_{\odot}$ the 
cluster metallicity. 
We add dependence of $\Gamma$ on metallicity, as indicated by
studies of LMXBs in globular cluster systems of elliptical galaxies
\citep{Kundu02, Jordan04}. 
The most detailed such study \citep{Jordan04} found a dependence
$\Gamma\propto \rho^{1.08\pm0.11} r_c^{2} (Z/Z_{\odot})^{0.33\pm0.1}$
for production of LMXBs in the clusters of M87.  

This analysis generally follows that of
\citet{Heinke03d}, differing in three respects: We study the effect of
metallicity in addition to central density and core radius (for most clusters we use the values from \citet{Harris96}, revision of 2003, otherwise using values from \citet{Heinke03d}). We update
numbers of likely qLMXBs and hard sources (bright and faint) for
several clusters, these being 
Terzan 5 (this paper), 
NGC 6266 (unpublished work by the authors; we use 5 qLMXBs, 7 hard sources with $L_X>10^{32}$ ergs s$^{-1}$, and 21 hard sources with $10^{32}>L_X>10^{31}$ ergs s$^{-1}$), 
47 Tuc \citep{Heinke05a}, and 
NGC 6397 \citep[changing the distance from 2.7 to 2.5 kpc, in accord with
][]{Gratton03}.  
We use the same numbers as \citet{Heinke03d} for the clusters
M80, M28, NGC 6752, $\omega$ Cen, M30, NGC 5904 (M5), M22, M13, NGC 6121 (M4), and NGC 6366.  
   Finally, we make two changes to the code computing the 
KS probabilities.  One change corrects a  
coding error in the program of \citet{Heinke03d}, which decreased the
best-fit densities for qLMXBs and bright CVs (see below).  
The other changes an assumption about the distribution of sources {\it within} 
each cluster.  The code distributes the sources along a fictional line 
segment, wherein each cluster occupies a length equal to its $\Gamma$, and 
the clusters are arranged in order of decreasing $\Gamma$.  The output is 
the K-S probability of finding such a distribution of sources along the line 
if the sources were distributed randomly (with an equal probability of 
source per unit $\Gamma$), thus a measure of the appropriateness of that 
choice of $\Gamma$.  The previous code distributed sources within a cluster 
evenly within its segment, artificially increasing the K-S probability of that 
choice of $\Gamma$.  This change randomizes the distribution of sources 
within the cluster segment, effectively decreasing all K-S probabilities.  

\begin{figure}
\figurenum{8}
\epsscale{2.0}
\plottwo{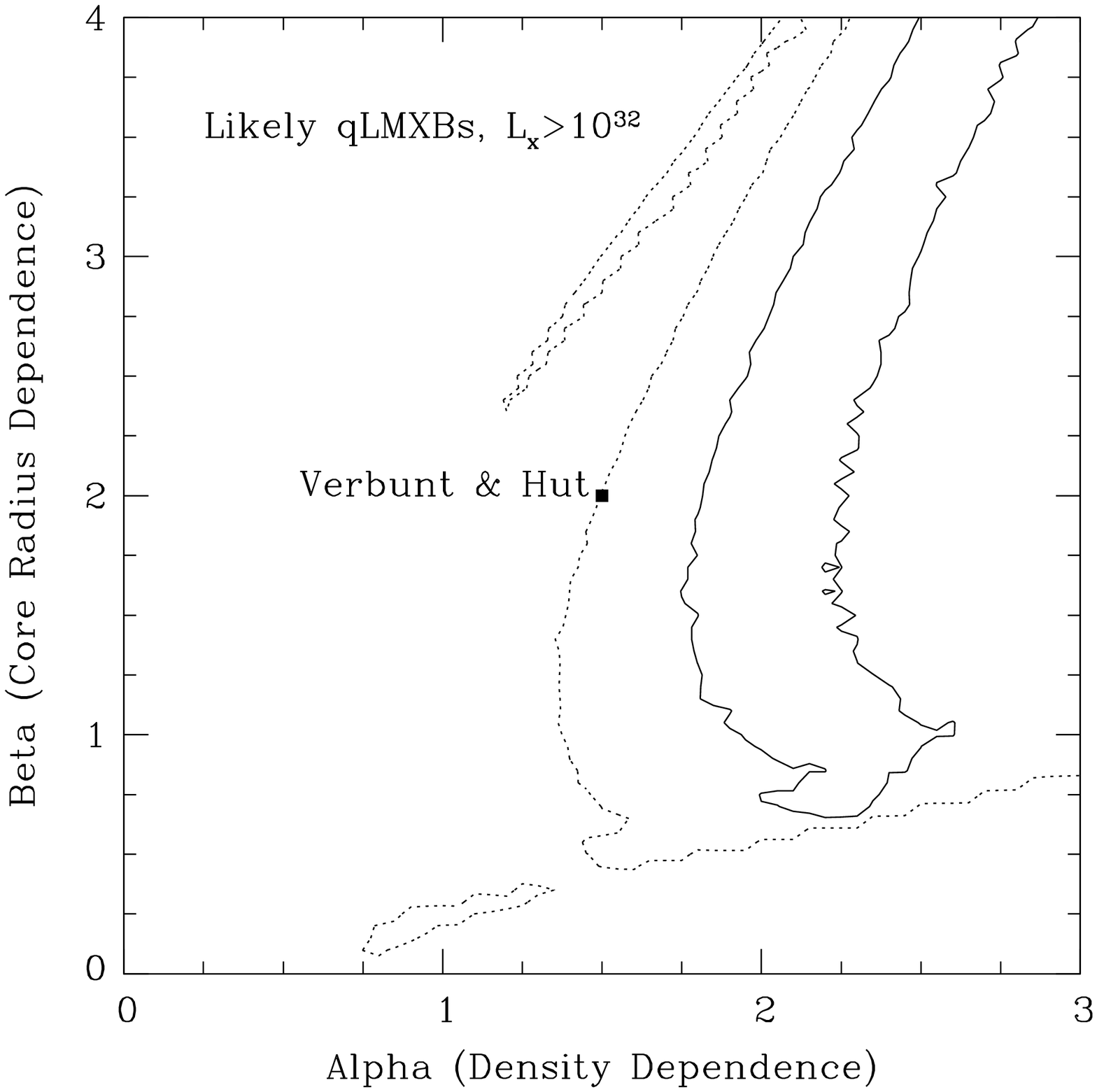}{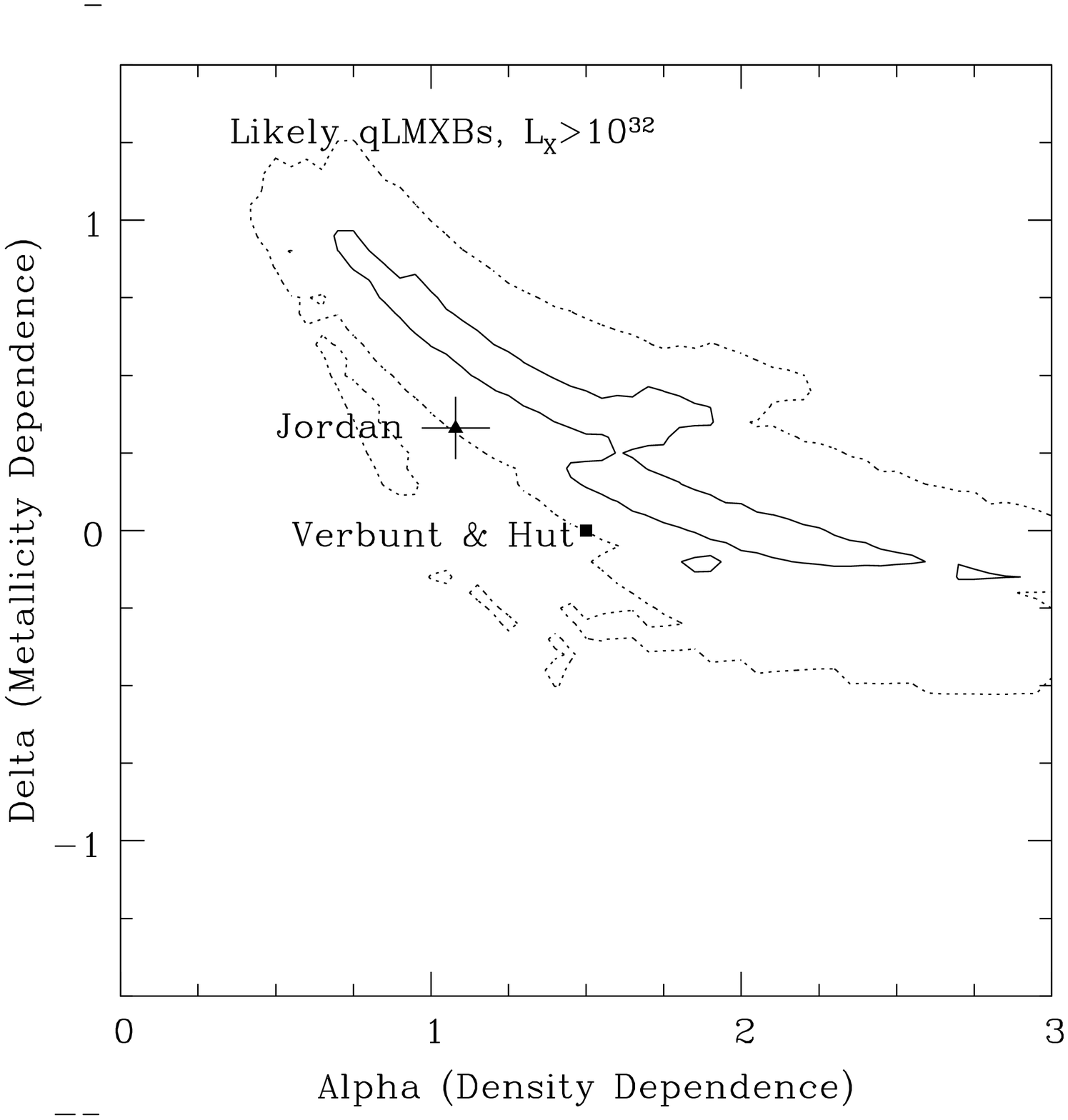}
\caption{Contours of K-S probability for different parametrizations of the 
production rate for likely qLMXBs.  Top, dependence of production rate on 
core density vs. core radius; bottom, dependence of production rate on core 
density vs. cluster metallicity.  Solid contours enclose $>$50\% KS 
probability; dotted contours enclose $>$10\% KS probability.
} \label{fig:qLMXBs}
\end{figure}

\begin{figure}
\figurenum{9}
\epsscale{2.0}
\plottwo{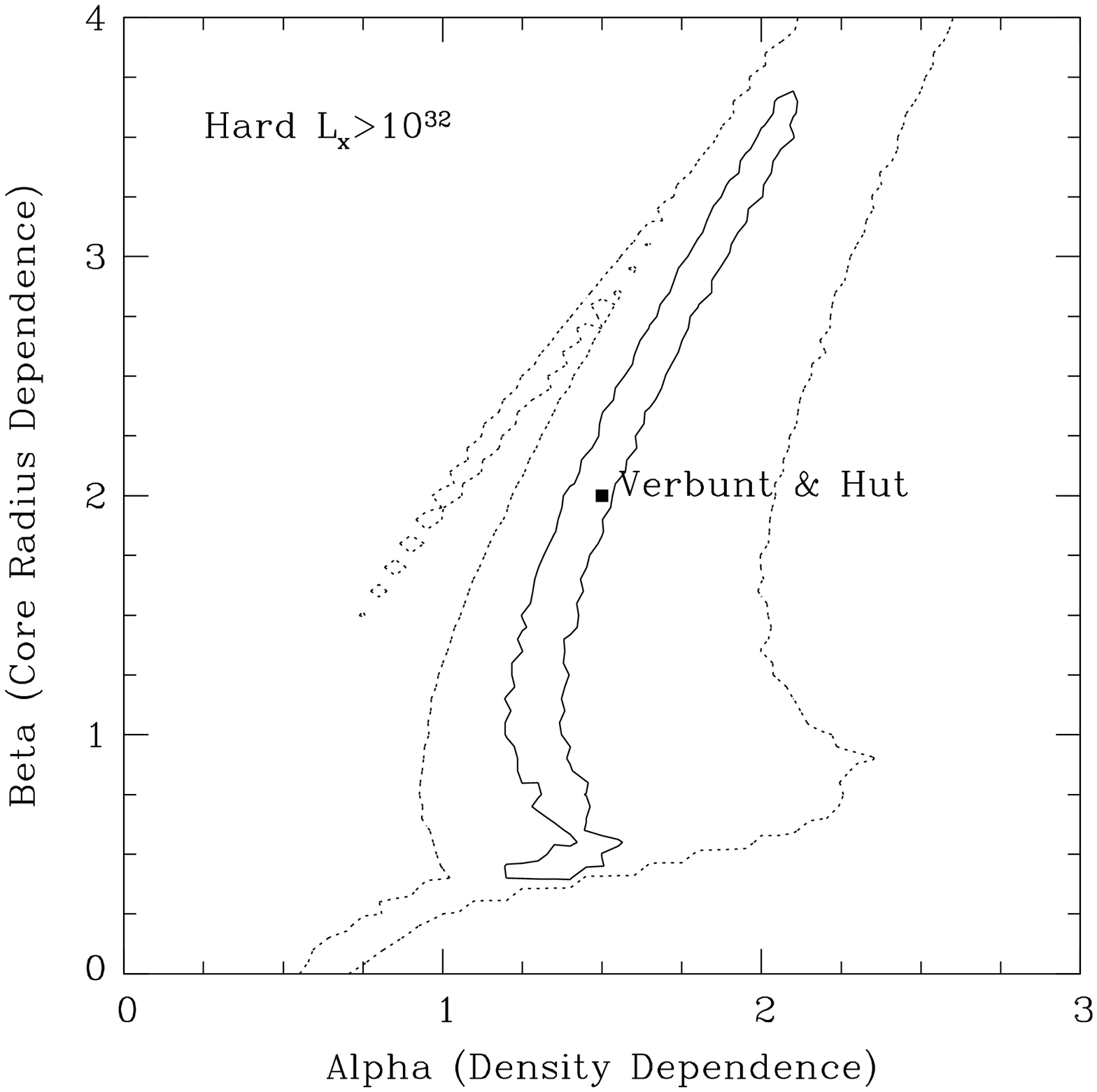}{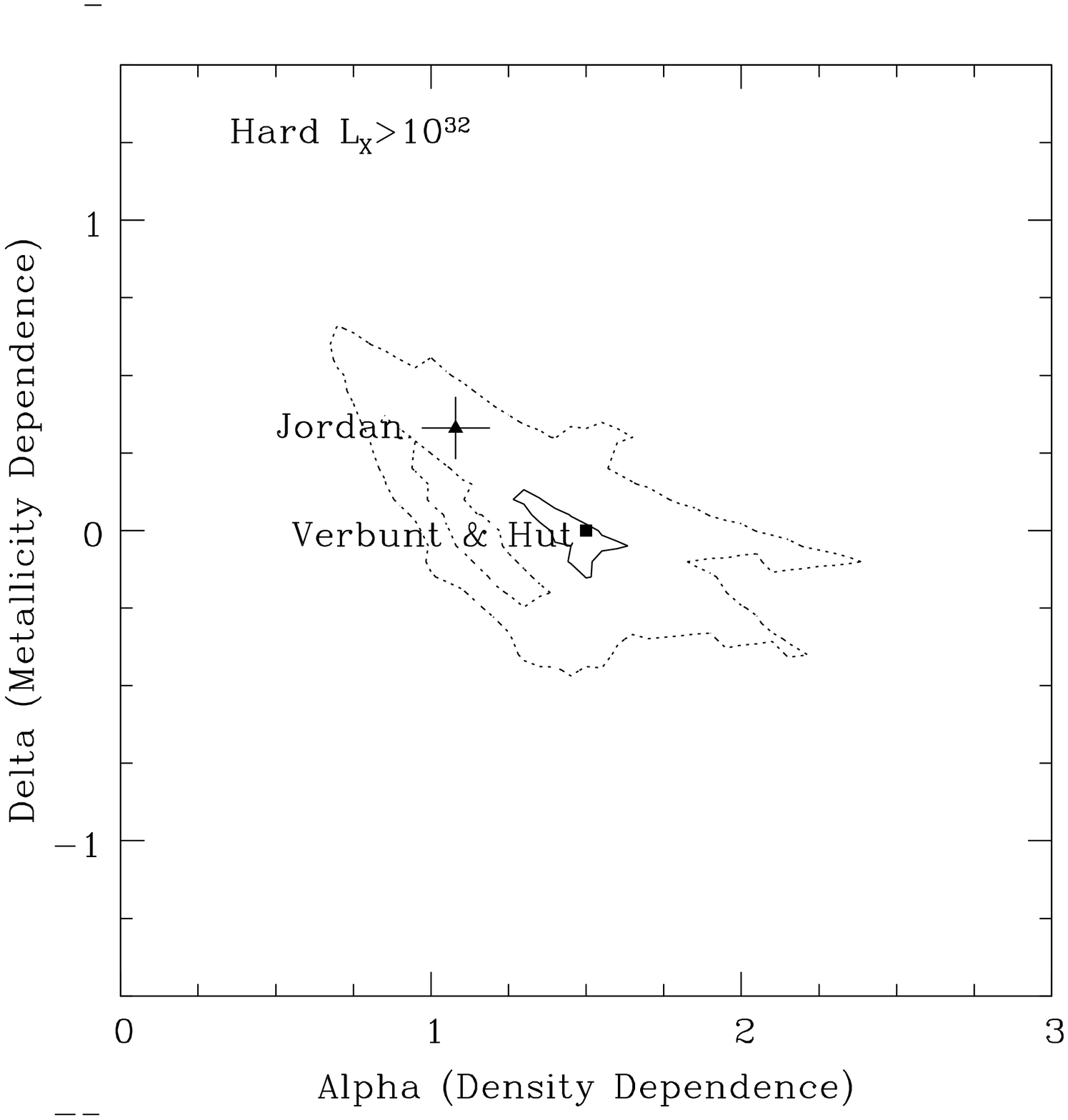}
\caption{ As in Figure \ref{fig:qLMXBs}, but for hard sources with 
$L_X>10^{32}$ ergs s$^{-1}$ (which may be dominated by bright CVs).
} \label{fig:bCVs}
\end{figure}

\begin{figure}
\figurenum{10}
\epsscale{2.0}
\plottwo{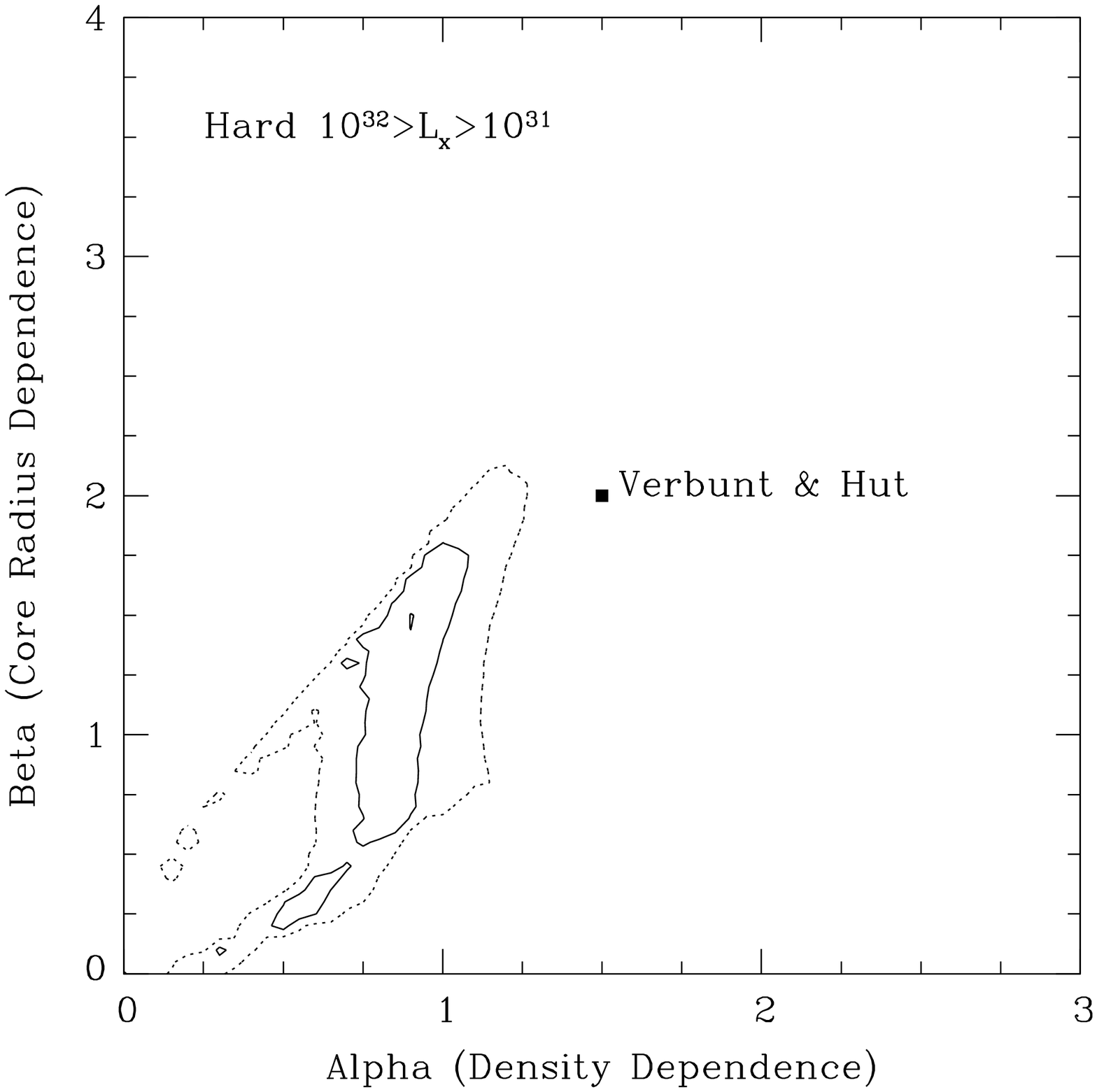}{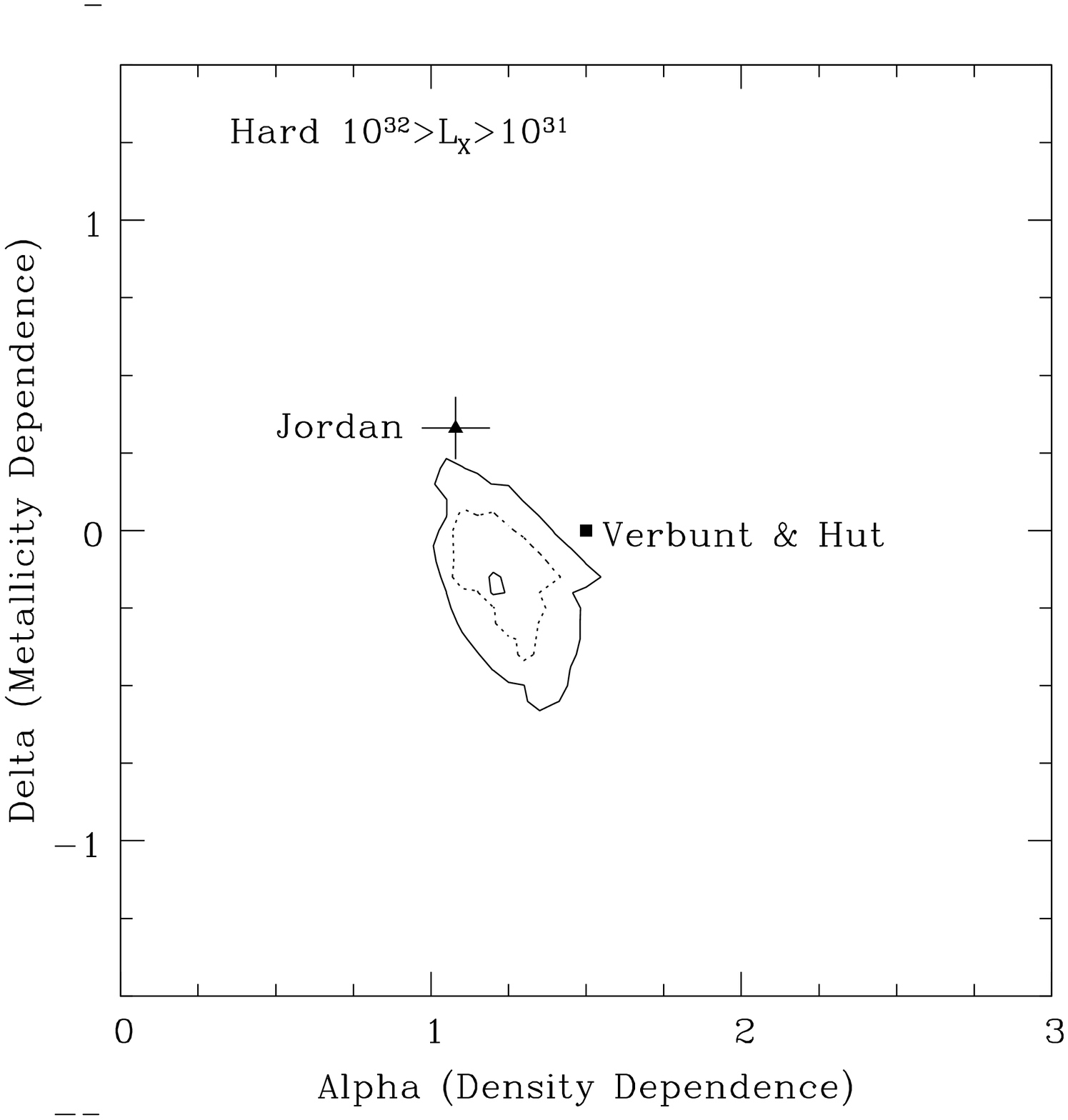}
\caption{As in Figure \ref{fig:qLMXBs}, but for hard sources with 
$10^{32}>L_X>10^{31}$ 
ergs s$^{-1}$ (which may be dominated by faint CVs). In the bottom panel 
(metallicity vs. density) we 
have also indicated a contour of 1\% KS probability with a solid line.
} \label{fig:fCVs}
\end{figure}

We first test a range of values for $\alpha$ and $\beta$ to explain
the distribution of each group of X-ray binaries: qLMXBs (limited to
those with $L_X$(0.5-2.5 keV)$>10^{32}$ ergs/s to reduce
incompleteness effects), bright hard sources ($L_X$(0.5-2.5
keV)$>10^{32}$ ergs/s), and faint hard sources ($L_X$(0.5-2.5
keV)$=10^{31}$--$10^{32}$ ergs/s).  For each choice of $\alpha$ and
$\beta$ we apply a Kolmogorov-Smirnov test \citep[implemented via 
][]{Press92}, comparing the distribution of observed sources with a 
distribution uniform in $\rho^{\alpha} r_c^{\beta}$.  We plot contours 
at 50\% and 10\% KS probabilities in the upper panels of figures 
\ref{fig:qLMXBs}, \ref{fig:bCVs} and \ref{fig:fCVs}.  

The likely qLMXBs are barely consistent with the Verbunt \& Hut 
predictions at the 10\% probability level.   
The hard bright sources are also consistent with the Verbunt \& Hut 
predictions, while the fainter hard sources are clearly inconsistent 
with the Verbunt \& Hut predictions, indicating a lower density 
dependence.  This result contradicts \citet{Heinke03d}, in that it finds a 
distinction between the distributions of bright hard sources and faint 
hard sources.  

To study the effect of metallicity, we assume $\beta=2$ (thus $\Gamma 
\propto r_c^2$) and test choices of $\alpha$ and $\delta$ (producing the 
lower panels of Figures \ref{fig:qLMXBs}, \ref{fig:bCVs}, and 
\ref{fig:fCVs}).  The densest clusters studied also 
tend to be the most metal-rich, so some degeneracy between density and 
metallicity is seen in these plots.  The likely qLMXBs are consistent 
at the 10\% confidence level with either the Verbunt \& Hut dependence 
on density and radius, or the Jordan et al. dependence.  
The brighter hard sources are also quite consistent with either of the 
suggested dependences.  However, the fainter hard sources are clearly 
inconsistent with either of the suggested dependencies.  In figure 
\ref{fig:fCVs} 
we have plotted, in addition to the contours of 10\% and 50\% KS probability, 
the 1\% KS probability.  Neither of the two suggested dependencies describe
 the observations of the faint hard sources, which require a lower 
dependence on either density or metallicity.

\section{Discussion}\label{s:disc}

We have not yet identified a clear metallicity dependence in the 
distribution of qLMXBs and bright CVs in globular clusters.  However, 
considering the evidence for a metallicity dependence in bright LMXBs in 
extragalactic globular clusters by \citet{Kundu02} and \citet{Jordan04}, 
we think it likely 
that additional observations of galactic globular clusters will show 
increasing evidence for a clear dependence.  From the 
information we have, it seems that fainter hard X-ray sources 
($10^{31}<L_X<10^{32}$ ergs s$^{-1}$) do not show a metallicity dependence 
of the strength observed for bright LMXBs by Jord\'{a}n et al.  

One important result from this study is the identification of a possible 
difference in the distribution among globular clusters of the brighter 
and fainter hard X-ray sources, above and below 
$L_X$(0.5-2.5 keV)$=10^{32}$ ergs s$^{-1}$.  The brighter sources are 
consistent with distributions $\propto \rho_c^{1.5} r_c^2$ or 
$\propto \rho_c^{1.08} r_c^2 (Z/Z_{\odot})^{0.33\pm0.1}$, as suggested for 
bright LMXBs by \citet{Verbunt87} and \citet{Jordan04}, while the 
fainter sources require a lesser dependence on density and/or  
metallicity.  Considering the looseness of the constraints on the 
brighter sources, it is still possible that the two groups arise from the same 
distribution; studies of additional clusters will allow this to be tested. 
Possible reasons for a difference include: The bright hard sources 
may contain a substantial number of qLMXBs and/or MSPs, with a different 
distribution than CVs \citep{Wijnands05a}.  The faint hard sources may 
include large numbers of active binaries, or primordial CVs, with a 
lower density distribution \citep[e.g., ][]{Bassa04}.  Finally, the 
densest clusters will produce many CVs and destroy them relatively quickly, 
due to the short timescale for their next interaction  
\citep[e.g.][]{Verbunt02}.  When formed (typically from turnoff stars at 
$\sim$0.8 \Msun), these CVs will be relatively brighter than after some 
Gyrs, so the densest clusters should have relatively more bright CVs.  
These effects can be seen also in the slope of the X-ray luminosity function
\citep{Pooley02b}, which flattens for the densest clusters.

\section{Conclusion}\label{s:concl}

Terzan 5 contains 28 X-ray sources above $L_X=10^{32}$ ergs 
s$^{-1}$ (0.5-2.5 keV), the richest population of X-ray sources so far observed 
in a globular cluster in this $L_X$ range.
  Twelve sources show soft X-ray colors suggesting a qLMXB nature.  
However, these sources are not generally well-fit by a simple 
hydrogen-atmosphere model, indicating that if these sources are qLMXBs, they 
have a substantial flux from harder nonthermal spectral components \citep[as 
seen in non-cluster systems, e.g. ][]{Rutledge01a}.  
Several faint X-ray sources 
have demonstrated substantial variability, up to a factor of five, between 
the 2000 and 2003 \Chandra\  observations.  

We constructed an X-ray color-color diagram for the sources in Terzan 5, 
for comparison with X-ray sources of similar luminosity at the Galactic 
center.  We find that the X-ray colors of the Galactic center sources are 
substantially harder than the relatively hard X-ray sources in Terzan 5, 
including controlling for the differences in photoelectric absorption.  
This suggests an intrinsic difference between the sources in Terzan 5 and the Galactic center.

Our study of the distribution of X-ray sources among globular clusters 
finds that likely qLMXBs, and hard X-ray sources with $L_X>10^{32}$ ergs 
s$^{-1}$ (which may be dominated by bright CVs) show consistency with 
the parametrizations by density, core radius, and metallicity of 
\citet{Verbunt87} and \citet{Jordan04}.  However, the hard X-ray sources 
with $10^{31}<L_X<10^{32}$ ergs s$^{-1}$ show a lesser dependence on 
density and metallicity.  

\acknowledgments

C.~O.~H. is supported by the Lindheimer Postdoctoral Fellowship at
Northwestern University.  We acknowledge very useful conversations
with Scott Ransom, Mike Muno, Ron Taam, and Jae Sub Hong, and the 
assistance of Patrick Broos and the Chandra X-ray Center. 

\bibliography{src_ref_list}

\begin{thebibliography}{69}
\expandafter\ifx\csname natexlab\endcsname\relax\def\natexlab#1{#1}\fi

\bibitem[{{Bassa} {et~al.}(2004){Bassa}, {Pooley}, {Homer}, \& {et
  al.}}]{Bassa04}
{Bassa}, C., {Pooley}, D., {Homer}, L.,  {et al.} 2004, \apj, 609, 755

\bibitem[{{Becker} {et~al.}(2003){Becker}, {Swartz}, {Pavlov}, {Elsner},
  {Grindlay}, {Mignani}, {Tennant}, {Backer}, {Pulone}, {Testa}, \&
  {Weisskopf}}]{Becker03}
{Becker}, W., {Swartz}, D.~A., {Pavlov}, G.~G., et al. 2003, \apj, 594, 798

\bibitem[{{Bellazzini} {et~al.}(1995){Bellazzini}, {Pasquali}, {Federici},
  {Ferraro}, \& {Pecci}}]{Bellazzini95}
{Bellazzini}, M., {Pasquali}, A., {Federici}, L., {Ferraro}, F.~R., \& {Pecci},
  F.~F. 1995, \apj, 439, 687

\bibitem[{{Bergbusch} \& {Van den Berg}(2001)}]{Bergbusch01}
{Bergbusch}, P.~A., \& {Van den Berg}, D.~A. 2001, \apj, 556, 322

\bibitem[{{Bogdanov} {et~al.}(2006){Bogdanov}, {Grindlay}, {Heinke}, {Camilo},
  {Freire}, \& {Becker}}]{Bogdanov06}
{Bogdanov}, S., {Grindlay}, J.~E., {Heinke}, C.~O., {Camilo}, F., {Freire},
  P.~C.~C., \& {Becker}, W. 2006, ApJ, in press; astro-ph/0604318

\bibitem[{{Bogdanov} {et~al.}(2005){Bogdanov}, {Grindlay}, \& {van den
  Berg}}]{Bogdanov05}
{Bogdanov}, S., {Grindlay}, J.~E., \& {van den Berg}, M. 2005, \apj, 630, 1029

\bibitem[{{Broos} {et~al.}(2002){Broos}, {Townsley}, {Getman}, \&
  {Bauer}}]{Broos02}
{Broos}, P., {Townsley}, L., {Getman}, K., \& {Bauer}, F. 2002, ACIS Extract,
  An ACIS Point Source Extraction Package, Pennsylvania State University,
  http://www.astro.psu.edu/xray/docs/TARA/ae\_users\_guide.html

\bibitem[{{Cackett} {et~al.}(2006){Cackett}, {Wijnands}, {Heinke}, {Pooley},
  {Lewin}, {Grindlay}, {Edmonds}, {Jonker}, \& {Miller}}]{Cackett06}
{Cackett}, E.~M., {Wijnands}, R., {Heinke}, C.~O., et al. 2006, \mnras, 369, 407

\bibitem[{{Campana} {et~al.}(1998){Campana}, {Colpi}, {Mereghetti}, {Stella},
  \& {Tavani}}]{Campana98a}
{Campana}, S., {Colpi}, M., {Mereghetti}, S., {Stella}, L., \& {Tavani}, M.
  1998, \aapr, 8, 279

\bibitem[{{Clark}(1975)}]{Clark75}
{Clark}, G.~W. 1975, \apjl, 199, L143

\bibitem[{{Cohn} {et~al.}(2002){Cohn}, {Lugger}, {Grindlay}, \&
  {Edmonds}}]{Cohn02}
{Cohn}, H.~N., {Lugger}, P.~M., {Grindlay}, J.~E., \& {Edmonds}, P.~D. 2002,
  \apj, 571, 818

\bibitem[{{Cool} {et~al.}(1995){Cool}, {Grindlay}, {Cohn}, {Lugger}, \&
  {Slavin}}]{Cool95}
{Cool}, A.~M., {Grindlay}, J.~E., {Cohn}, H.~N., {Lugger}, P.~M., \& {Slavin},
  S.~D. 1995, \apj, 439, 695

\bibitem[{{Cool} {et~al.}(2002){Cool}, {Haggard}, \& {Carlin}}]{Cool02b}
{Cool}, A.~M., {Haggard}, D., \& {Carlin}, J.~L. 2002, in ASP Conf. Ser. 265:
  Omega Centauri, A Unique Window into Astrophysics, 277

\bibitem[{{Damiani} {et~al.}(1997){Damiani}, {Maggio}, {Micela}, \&
  {Sciortino}}]{Damiani97}
{Damiani}, F., {Maggio}, A., {Micela}, G., \& {Sciortino}, S. 1997, \apj, 483,
  370

\bibitem[{{Edmonds} {et~al.}(2003){Edmonds}, {Gilliland}, {Heinke}, \&
  {Grindlay}}]{Edmonds03a}
{Edmonds}, P.~D., {Gilliland}, R.~L., {Heinke}, C.~O., \& {Grindlay}, J.~E.
  2003, \apj, 596, 1177

\bibitem[{{Edmonds} {et~al.}(1999){Edmonds}, {Grindlay}, {Cool}, {Cohn},
  {Lugger}, \& {Bailyn}}]{Edmonds99}
{Edmonds}, P.~D., {Grindlay}, J.~E., {Cool}, A., {Cohn}, H., {Lugger}, P., \&
  {Bailyn}, C. 1999, \apj, 516, 250

\bibitem[{{Feigelson} {et~al.}(2002){Feigelson}, {Broos}, {Gaffney}, {Garmire},
  {Hillenbrand}, {Pravdo}, {Townsley}, \& {Tsuboi}}]{Feigelson02}
{Feigelson}, E.~D., {Broos}, P., {Gaffney}, J.~A., et al. 2002,
  \apj, 574, 258

\bibitem[{{Freeman} {et~al.}(2002){Freeman}, {Kashyap}, {Rosner}, \&
  {Lamb}}]{Freeman02}
{Freeman}, P.~E., {Kashyap}, V., {Rosner}, R., \& {Lamb}, D.~Q. 2002, \apjs,
  138, 185

\bibitem[{{Fruchter} \& {Goss}(2000)}]{Fruchter00}
{Fruchter}, A.~S., \& {Goss}, W.~M. 2000, \apj, 536, 865

\bibitem[{{Gendre} {et~al.}(2003){Gendre}, {Barret}, \& {Webb}}]{Gendre03a}
{Gendre}, B., {Barret}, D., \& {Webb}, N.~A. 2003, \aap, 400, 521

\bibitem[{{Giacconi} {et~al.}(2001){Giacconi}, {Rosati}, {Tozzi}, \& {et
  al.}}]{Giacconi01}
{Giacconi}, R., {Rosati}, P., {Tozzi}, P., {et al.} 2001, \apj, 551, 624

\bibitem[{{Gratton} {et~al.}(2003){Gratton}, {Bragaglia}, {Carretta}, \& {et
  al.}}]{Gratton03}
{Gratton}, R.~G., {Bragaglia}, A., {Carretta}, E., {et al.} 2003, \aap, 408,
  529

\bibitem[{{Grindlay}(1987)}]{Grindlay87}
{Grindlay}, J.~E. 1987, in IAU Symp. 125: The Origin and Evolution of Neutron
  Stars, ed. D.~J. {Helfand} \& J.-H. {Huang}, 173--184

\bibitem[{{Grindlay} {et~al.}(2002){Grindlay}, {Camilo}, {Heinke}, {Edmonds},
  {Cohn}, \& {Lugger}}]{Grindlay02}
{Grindlay}, J.~E., {Camilo}, F., {Heinke}, C.~O., {Edmonds}, P.~D., {Cohn}, H.,
  \& {Lugger}, P. 2002, \apj, 581, 470

\bibitem[{{Grindlay} {et~al.}(1995){Grindlay}, {Cool}, {Callanan}, {Bailyn},
  {Cohn}, \& {Lugger}}]{Grindlay95}
{Grindlay}, J.~E., {Cool}, A.~M., {Callanan}, P.~J., {Bailyn}, C.~D., {Cohn},
  H.~N., \& {Lugger}, P.~M. 1995, \apjl, 455, L47

\bibitem[{{Grindlay} {et~al.}(2001{\natexlab{a}}){Grindlay}, {Heinke},
  {Edmonds}, \& {Murray}}]{Grindlay01a}
{Grindlay}, J.~E., {Heinke}, C., {Edmonds}, P.~D., \& {Murray}, S.~S.
  2001{\natexlab{a}}, Science, 292, 2290

\bibitem[{{Grindlay} {et~al.}(2001{\natexlab{b}}){Grindlay}, {Heinke},
  {Edmonds}, {Murray}, \& {Cool}}]{Grindlay01b}
{Grindlay}, J.~E., {Heinke}, C.~O., {Edmonds}, P.~D., {Murray}, S.~S., \&
  {Cool}, A.~M. 2001{\natexlab{b}}, \apjl, 563, L53

\bibitem[{{Grindlay} {et~al.}(1984){Grindlay}, {Hertz}, {Steiner}, {Murray}, \&
  {Lightman}}]{Grindlay84}
{Grindlay}, J.~E., {Hertz}, P., {Steiner}, J.~E., {Murray}, S.~S., \&
  {Lightman}, A.~P. 1984, \apjl, 282, L13

\bibitem[{{Hambly} {et~al.}(2001){Hambly}, {MacGillivray}, {Read}, {Tritton},
  {Thomson}, {Kelly}, {Morgan}, {Smith}, {Driver}, {Williamson}, {Parker},
  {Hawkins}, {Williams}, \& {Lawrence}}]{Hambly01}
{Hambly}, N.~C., {MacGillivray}, H.~T., {Read}, M.~A., et al. 2001, \mnras, 326, 1279

\bibitem[{{Harris}(1996)}]{Harris96}
{Harris}, W.~E. 1996, \aj, 112, 1487

\bibitem[{{Heinke} {et~al.}(2003{\natexlab{a}}){Heinke}, {Edmonds}, {Grindlay},
  {Lloyd}, {Cohn}, \& {Lugger}}]{Heinke03b}
{Heinke}, C.~O., {Edmonds}, P.~D., {Grindlay}, J.~E., {Lloyd}, D.~A., {Cohn},
  H.~N., \& {Lugger}, P.~M. 2003{\natexlab{a}}, \apj, 590, 809

\bibitem[{{Heinke} {et~al.}(2005{\natexlab{a}}){Heinke}, {Grindlay}, \&
  {Edmonds}}]{Heinke05b}
{Heinke}, C.~O., {Grindlay}, J.~E., \& {Edmonds}, P.~D. 2005{\natexlab{a}},
  \apj, 622, 556

\bibitem[{{Heinke} {et~al.}(2005{\natexlab{b}}){Heinke}, {Grindlay}, {Edmonds},
  {Cohn}, {Lugger}, {Camilo}, {Bogdanov}, \& {Freire}}]{Heinke05a}
{Heinke}, C.~O., {Grindlay}, J.~E., {Edmonds}, P.~D., et al. 2005{\natexlab{b}},
  \apj, 625, 796

\bibitem[{{Heinke} {et~al.}(2003{\natexlab{b}}){Heinke}, {Grindlay}, {Lloyd},
  \& {Edmonds}}]{Heinke03a}
{Heinke}, C.~O., {Grindlay}, J.~E., {Lloyd}, D.~A., \& {Edmonds}, P.~D.
  2003{\natexlab{b}}, \apj, 588, 452

\bibitem[{{Heinke} {et~al.}(2003{\natexlab{c}}){Heinke}, {Grindlay}, {Lugger},
  {Cohn}, {Edmonds}, {Lloyd}, \& {Cool}}]{Heinke03d}
{Heinke}, C.~O., {Grindlay}, J.~E., {Lugger}, P.~M., {Cohn}, H.~N., {Edmonds},
  P.~D., {Lloyd}, D.~A., \& {Cool}, A.~M. 2003{\natexlab{c}}, \apj, 598, 501

\bibitem[{{Heinke} {et~al.}(2006){Heinke}, {Rybicki}, {Narayan}, \&
  {Grindlay}}]{Rybicki05}
{Heinke}, C.~O., {Rybicki}, G.~B., {Narayan}, R., \& {Grindlay}, J.~E. 2006,
  ApJ (in press; astro-ph/0506563), 644, 1090

\bibitem[{{Hessels} {et~al.}(2006){Hessels}, {Ransom}, {Stairs}, {Freire},
  {Kaspi}, \& {Camilo}}]{Hessels06}
{Hessels}, J.~W.~T., {Ransom}, S.~M., {Stairs}, I.~H., {Freire}, P.~C.~C.,
  {Kaspi}, V.~M., \& {Camilo}, F. 2006, Science, 311, 1901

\bibitem[{{in't Zand} {et~al.}(2001){in't Zand}, {van Kerkwijk}, {Pooley},
  {Verbunt}, {Wijnands}, \& {Lewin}}]{intZand01}
{in't Zand}, J.~J.~M., {van Kerkwijk}, M.~H., {Pooley}, D., {Verbunt}, F.,
  {Wijnands}, R., \& {Lewin}, W.~H.~G. 2001, \apjl, 563, L41

\bibitem[{{Ivanova}(2006)}]{Ivanova06}
{Ivanova}, N. 2006, \apj, 636, 979

\bibitem[{{Ivanova} {et~al.}(2005){Ivanova}, {Fregeau}, \& {Rasio}}]{Ivanova04}
{Ivanova}, N., {Fregeau}, J.~M., \& {Rasio}, F.~A. 2005, in ASP Conf. Ser. 328:
  Binary Radio Pulsars, ed. F.~A. {Rasio} \& I.~H. {Stairs}, 231--+

\bibitem[{{Johnston} \& {Verbunt}(1996)}]{Johnston96}
{Johnston}, H.~M., \& {Verbunt}, F. 1996, \aap, 312, 80

\bibitem[{{Jonker} {et~al.}(2004){Jonker}, {Galloway}, {McClintock}, {Buxton},
  {Garcia}, \& {Murray}}]{Jonker04}
{Jonker}, P.~G., {Galloway}, D.~K., {McClintock}, J.~E., {Buxton}, M.,
  {Garcia}, M., \& {Murray}, S. 2004, \mnras, 354, 666

\bibitem[{{Jord{\'a}n} {et~al.}(2004){Jord{\'a}n}, {C{\^o}t{\'e}}, {Ferrarese},
  {Blakeslee}, {Mei}, {Merritt}, {Milosavljevi{\'c}}, {Peng}, {Tonry}, \&
  {West}}]{Jordan04}
{Jord{\'a}n}, A., {C{\^o}t{\'e}}, P., {Ferrarese}, L., et al. 2004, \apj, 613, 279

\bibitem[{{Kong} {et~al.}(2006){Kong}, {Bassa}, {Pooley}, {Lewin}, {Homer},
  {Verbunt}, {Anderson}, \& {Margon}}]{Kong06}
{Kong}, A.~K.~H., {Bassa}, C., {Pooley}, D., et al. 2006, ApJ, submitted, astro-ph/0603374

\bibitem[{{Kundu} {et~al.}(2002){Kundu}, {Maccarone}, \& {Zepf}}]{Kundu02}
{Kundu}, A., {Maccarone}, T.~J., \& {Zepf}, S.~E. 2002, \apjl, 574, L5

\bibitem[{{Kundu} {et~al.}(2003){Kundu}, {Maccarone}, {Zepf}, \&
  {Puzia}}]{Kundu03}
{Kundu}, A., {Maccarone}, T.~J., {Zepf}, S.~E., \& {Puzia}, T.~H. 2003, \apjl,
  589, L81

\bibitem[{{Liedahl} {et~al.}(1995){Liedahl}, {Osterheld}, \&
  {Goldstein}}]{Liedahl95}
{Liedahl}, D.~A., {Osterheld}, A.~L., \& {Goldstein}, W.~H. 1995, \apjl, 438,
  L115

\bibitem[{{Lugger} {et~al.}(2006){Lugger}, {Cohn}, {Heinke}, {Grindlay}, \&
  {Edmonds}}]{Lugger06}
{Lugger}, P.~M., {Cohn}, H.~N., {Heinke}, C.~O., {Grindlay}, J.~E., \&
  {Edmonds}, P.~D. 2006, \apj, submitted

\bibitem[{{Maccarone} {et~al.}(2003){Maccarone}, {Kundu}, \&
  {Zepf}}]{Maccarone03}
{Maccarone}, T.~J., {Kundu}, A., \& {Zepf}, S.~E. 2003, \apj, 586, 814

\bibitem[{{Maccarone} {et~al.}(2004){Maccarone}, {Kundu}, \&
  {Zepf}}]{Maccarone04a}
---. 2004, \apj, 606, 430

\bibitem[{{Makishima} {et~al.}(1981){Makishima}, {Ohashi}, {Inoue}, \& {et
  al.}}]{Makishima81}
{Makishima}, K., {Ohashi}, T., {Inoue}, H., {et al.} 1981, \apjl, 247, L23

\bibitem[{{Monet} {et~al.}(2003){Monet}, {Levine}, {Canzian}, \&
  {et~al.}}]{Monet03}
{Monet}, D.~G., {Levine}, S.~E., {Canzian}, B.,  {et~al.} 2003, \aj, 125, 984

\bibitem[{{Muno} {et~al.}(2003){Muno}, {Baganoff}, {Bautz}, \& {et
  al.}}]{Muno03}
{Muno}, M.~P., {Baganoff}, F.~K., {Bautz}, M.~W., {et al.} 2003, \apj, 589,
  225

\bibitem[{{Origlia} \& {Rich}(2004)}]{Origlia04}
{Origlia}, L., \& {Rich}, R.~M. 2004, \aj, 127, 3422

\bibitem[{{Patterson} \& {Raymond}(1985)}]{Patterson85}
{Patterson}, J., \& {Raymond}, J.~C. 1985, \apj, 292, 535

\bibitem[{{Pooley} \& {Hut}(2006)}]{Pooley06a}
{Pooley}, D., \& {Hut}, P. 2006, ApJ, submitted; astro-ph/0605048

\bibitem[{{Pooley} {et~al.}(2003){Pooley}, {Lewin}, {Anderson}, \& {et
  al.}}]{Pooley03}
{Pooley}, D., {Lewin}, W.~H.~G., {Anderson}, S.~F., {et al.} 2003, \apjl,
  591, L131

\bibitem[{{Pooley} {et~al.}(2002{\natexlab{a}}){Pooley}, {Lewin}, {Homer}, \&
  {et al.}}]{Pooley02a}
{Pooley}, D., {Lewin}, W.~H.~G., {Homer}, L., {et al.} 2002{\natexlab{a}},
  \apj, 569, 405

\bibitem[{{Pooley} {et~al.}(2002{\natexlab{b}}){Pooley}, {Lewin}, {Verbunt}, \&
  {et al.}}]{Pooley02b}
{Pooley}, D., {Lewin}, W.~H.~G., {Verbunt}, F., {et al.} 2002{\natexlab{b}},
  \apj, 573, 184

\bibitem[{{Predehl} {et~al.}(2003){Predehl}, {Costantini}, {Hasinger}, \&
  {Tanaka}}]{Predehl03}
{Predehl}, P., {Costantini}, E., {Hasinger}, G., \& {Tanaka}, Y. 2003,
  Astronomische Nachrichten, 324, 73

\bibitem[{{Press} {et~al.}(1992){Press}, {Teukolsky}, {Vetterling}, \&
  {Flannery}}]{Press92}
{Press}, W.~H., {Teukolsky}, S.~A., {Vetterling}, W.~T., \& {Flannery}, B.~P.
  1992, {Numerical recipes in FORTRAN. The art of scientific computing}
  (Cambridge: University Press, |c1992, 2nd ed.)

\bibitem[{{Ransom} {et~al.}(2005){Ransom}, {Hessels}, {Stairs}, {Freire},
  {Camilo}, {Kaspi}, \& {Kaplan}}]{Ransom05}
{Ransom}, S.~M., {Hessels}, J.~W.~T., {Stairs}, I.~H., {Freire}, P.~C.~C.,
  {Camilo}, F., {Kaspi}, V.~M., \& {Kaplan}, D.~L. 2005, Science, 307, 892

\bibitem[{{Rutledge} {et~al.}(2001){Rutledge}, {Bildsten}, {Brown}, {Pavlov},
  \& {Zavlin}}]{Rutledge01a}
{Rutledge}, R.~E., {Bildsten}, L., {Brown}, E.~F., {Pavlov}, G.~G., \&
  {Zavlin}, V.~E. 2001, \apj, 551, 921

\bibitem[{{Rutledge} {et~al.}(2002){Rutledge}, {Bildsten}, {Brown}, {Pavlov},
  \& {Zavlin}}]{Rutledge02a}
---. 2002, \apj, 578, 405

\bibitem[{{Sarazin} {et~al.}(2003){Sarazin}, {Kundu}, {Irwin}, {Sivakoff},
  {Blanton}, \& {Randall}}]{Sarazin03}
{Sarazin}, C.~L., {Kundu}, A., {Irwin}, J.~A., {Sivakoff}, G.~R., {Blanton},
  E.~L., \& {Randall}, S.~W. 2003, \apj, 595, 743

\bibitem[{{Verbunt}(2003)}]{Verbunt02}
{Verbunt}, F. 2003, in ASP Conf. Ser. 296: New Horizons in Globular Cluster
  Astronomy, 245, astro--ph/0210057

\bibitem[{{Verbunt} \& {Hut}(1987)}]{Verbunt87}
{Verbunt}, F., \& {Hut}, P. 1987, in IAU Symp. 125: The Origin and Evolution of
  Neutron Stars, 187

\bibitem[{{Verbunt} \& {Lewin}(2006)}]{Verbunt04}
{Verbunt}, F., \& {Lewin}, W.~H.~G. 2006, in Compact Stellar X-ray Sources,
  eds. W.~H.~G.~Lewin \& M.~van der Klis, astro--ph/0404136

\bibitem[{{Wijnands} {et~al.}(2005){Wijnands}, {Heinke}, {Pooley}, {Edmonds},
  {Lewin}, {Grindlay}, {Jonker}, \& {Miller}}]{Wijnands05a}
{Wijnands}, R., {Heinke}, C.~O., {Pooley}, D., et al. 2005, \apj,
  618, 883

\end{thebibliography}
\bibliographystyle{apj}

\clearpage

\LongTables

\begin{deluxetable}{lccccccccr}
\tablecolumns{10}
\tabletypesize{\footnotesize}
\tablewidth{7.5truein}
\tablecaption{\textbf{X-ray Sources in Terzan 5}}
\tablehead{
\multicolumn{2}{c}{\textbf{Source}} & \colhead{RA} &
\colhead{Dec} & \colhead{Distance} &
\multicolumn{2}{c}{Counts} & \multicolumn{2}{c}{$L_X$, ergs s$^{-1}$}
& \colhead{Notes} \\
 (CX) & (CXOGLB J)  & (HH:MM:SS Err) & (DD:MM:SS Err) & (\arcsec)  &
(0.5-2.0) & (2-6) & (1-6) & (0.5-2.5) & \\
}
\startdata
1  & 174804.5-244641 & 17:48:04.587  0.001 & -24:46:41.95  0.02 & 5.41 & $95.4^{+11.5}_{-9.7}$ & $267.3^{+18.6}_{-16.3}$ & $2027^{+132}_{-114}$ & $2290^{+1020}_{-670}$ &  Y \\
2  & 174805.4-244637 & 17:48:05.413  0.001 & -24:46:37.67  0.02 & 10.20 & $184.7^{+15.3}_{-13.6}$ & $128.6^{+13.6}_{-11.3}$ & $1696^{+120}_{-103}$ & $2840^{+1020}_{-680}$ &  W3,q,Y \\ 
3  & 174805.2-244647 & 17:48:05.236  0.002 & -24:46:47.38  0.02 & 5.00 & $98.6^{+11.6}_{-9.9}$ & $141.6^{+14.2}_{-11.8}$ & $1328^{+109}_{-91}$ & $950^{+660}_{-240}$ &  L,Y \\ 
4  & 174804.7-244709 & 17:48:04.712  0.002 & -24:47:09.06  0.02 & 24.04 & $67.8^{+9.9}_{-8.2}$ & $135.7^{+14.0}_{-11.6}$ & $1137^{+102}_{-84}$ & $793^{+661}_{-237}$ &  Y \\ 
5  & 174802.6-244602 & 17:48:02.663  0.002 & -24:46:02.45  0.03 & 52.53 & $62.8^{+9.7}_{-7.9}$ & $130.8^{+13.7}_{-11.4}$ & $1033^{+97}_{-79}$ & $1500^{+900}_{-540}$ &  \\ 
6  & 174804.4-244638 & 17:48:04.427  0.002 & -24:46:38.21  0.03 & 9.55 & $45.5^{+8.5}_{-6.7}$ & $84.4^{+11.6}_{-9.1}$ & $714^{+85}_{-66}$ & $540^{+650}_{-230}$ &  W5,V?,Y \\ 
7  & 174804.1-244640 & 17:48:04.108  0.002 & -24:46:40.40  0.03 & 11.86 & $14.6^{+5.7}_{-3.7}$ & $91.4^{+11.9}_{-9.5}$ & $608^{+79}_{-62}$ & $160^{+39}_{-28}$ &  W9,Y \\
8  & 174804.3-244703 & 17:48:04.398  0.002 & -24:47:03.60  0.03 & 19.68 & $27.7^{+7.1}_{-5.2}$ & $76.7^{+11.1}_{-8.7}$ & $631^{+86}_{-66}$ & $450^{+690}_{-240}$ & W6,V? \\ 
9  & 174804.8-244644 & 17:48:04.823  0.003 & -24:46:44.81  0.03 & 1.16 & $61.5^{+9.6}_{-7.8}$ & $34.5^{+8.3}_{-5.8}$ & $491^{+69}_{-53}$ & $930^{+750}_{-360}$ &  W4,q? \\ 
10  & 174805.0-244641 & 17:48:05.048  0.003 & -24:46:41.10  0.03 & 4.51 & $21.7^{+6.5}_{-4.6}$ & $63.5^{+10.4}_{-7.9}$ & $479^{+74}_{-54}$ & $154^{+39}_{-28}$ &  MSP \\ 
11  & 174804.2-244642 & 17:48:04.248  0.003 & -24:46:42.20  0.03 & 9.42 & $18.6^{+6.2}_{-4.2}$ & $65.4^{+10.5}_{-8.0}$ & $480^{+74}_{-54}$ & $177^{+42}_{-31}$ &  W7 \\ 
12  & 174806.2-244642 & 17:48:06.210  0.003 & -24:46:42.63  0.04 & 17.94 & $50.8^{+8.8}_{-7.1}$ & $28.7^{+7.8}_{-5.2}$ & $434^{+69}_{-51}$ & $358^{+57}_{-47}$ & W2,q? \\ 
13  & 174803.8-244641 & 17:48:03.856  0.003 & -24:46:41.55  0.04 & 14.73 & $5.7^{+4.5}_{-2.2}$ & $58.5^{+10.1}_{-7.5}$ & $377^{+68}_{-48}$ & $79^{+32}_{-21}$ &  V? \\
14  & 174805.3-244652 & 17:48:05.383  0.002 & -24:46:52.75  0.03 & 9.98 & $34.8^{+7.7}_{-5.8}$ & $27.8^{+7.8}_{-5.1}$ & $365^{+73}_{-49}$ & $541^{+773}_{-275}$ & V?,q? \\ 
15  & 174804.2-244648 & 17:48:04.203  0.004 & -24:46:48.00  0.05 & 9.99 & $41.7^{+8.2}_{-6.4}$ & $18.4^{+6.8}_{-4.1}$ & $339^{+62}_{-45}$ & $271^{+52}_{-41}$ &  W8,q? \\ 
16  & 174803.5-244649 & 17:48:03.579  0.003 & -24:46:49.54  0.04 & 18.59 & $22.7^{+6.6}_{-4.7}$ & $34.6^{+8.4}_{-5.7}$ & $297^{+61}_{-40}$ & $143^{+39}_{-27}$ &  W10,V? \\ 
17  & 174804.3-244636 & 17:48:04.345  0.003 & -24:46:36.02  0.05 & 11.92 & $11.6^{+5.4}_{-3.3}$ & $42.4^{+8.9}_{-6.4}$ & $321^{+66}_{-46}$ & $330^{+654}_{-227}$ & \\
18  & 174805.2-244651 & 17:48:05.271  0.003 & -24:46:51.38  0.04 & 7.93 & $36.8^{+7.9}_{-6.0}$ & $16.8^{+6.7}_{-3.9}$ & $373^{+76}_{-53}$ & $640^{+813}_{-289}$ &  q? \\ 
19  & 174804.6-244645 & 17:48:04.625  0.004 & -24:46:45.32  0.05 & 3.83 & $14.4^{+5.8}_{-3.7}$ & $33.4^{+8.2}_{-5.6}$ & $261^{+59}_{-39}$ & $351^{+650}_{-220}$ &  V \\ 
20  & 174803.0-244640 & 17:48:03.064  0.004 & -24:46:40.92  0.05 & 25.45 & $29.7^{+7.4}_{-5.3}$ & $9.7^{+5.8}_{-2.9}$ & $189^{+50}_{-35}$ & $2700^{+1200}_{-800}$ &  s?,q?,V,Y \\ 
21  & 174804.2-244625 & 17:48:04.285  0.003 & -24:46:25.47  0.04 & 21.48 & $24.9^{+6.8}_{-4.9}$ & $10.9^{+6.0}_{-3.1}$ & $265^{+63}_{-46}$ & $220^{+57}_{-43}$ &  q? \\
22  & 174806.1-244617 & 17:48:06.188  0.005 & -24:46:17.58  0.07 & 32.66 & $8.8^{+5.0}_{-2.8}$ & $26.7^{+7.7}_{-5.0}$ & $209^{+57}_{-36}$ & $61^{+30}_{-17}$ &  \\
23  & 174803.5-244646 & 17:48:03.540  0.005 & -24:46:46.02  0.05 & 18.65 & $7.7^{+4.8}_{-2.6}$ & $23.6^{+7.4}_{-4.7}$ & $176^{+52}_{-32}$ & $43^{+27}_{-14}$ &  \\ 
24  & 174805.1-244645 & 17:48:05.105  0.005 & -24:46:45.90  0.09 & 2.81 & $14.7^{+5.7}_{-3.7}$ & $14.6^{+6.5}_{-3.6}$ & $170^{+52}_{-32}$ & $102^{+36}_{-24}$ &  \\ 
25  & 174804.8-244648 & 17:48:04.831  0.004 & -24:46:48.87  0.06 & 3.83 & $24.8^{+6.8}_{-4.9}$ & $3.9^{+4.9}_{-1.8}$ & $178^{+50}_{-35}$ & $170^{+46}_{-34}$ &  q? \\ 
26  & 174803.8-244645 & 17:48:03.869  0.003 & -24:46:45.92  0.07 & 14.15 & $8.7^{+5.0}_{-2.8}$ & $19.5^{+7.0}_{-4.2}$ & $159^{+50}_{-30}$ & $65^{+31}_{-18}$ &  \\
27  & 174806.1-244624 & 17:48:06.107  0.005 & -24:46:24.16  0.06 & 26.64 & $22.8^{+6.7}_{-4.7}$ & $1.9^{+4.6}_{-1.1}$ & $104^{+36}_{-23}$ & $653^{+751}_{-361}$ &  q? \\ 
28  & 174804.6-244648 & 17:48:04.698  0.005 & -24:46:48.30  0.05 & 4.23 & $1.9^{+3.7}_{-1.2}$ & $21.7^{+7.2}_{-4.5}$ & $151^{+51}_{-32}$ & $28^{+19}_{-11}$ &  \\
29  & 174804.7-244642 & 17:48:04.748  0.006 & -24:46:42.66  0.06 & 3.31 & $18.7^{+6.2}_{-4.2}$ & $2.8^{+4.8}_{-1.5}$ & $129^{+43}_{-29}$ & $422^{+726}_{-255}$ &  q? \\ 
30  & 174804.5-244640 & 17:48:04.591  0.005 & -24:46:40.37  0.05 & 6.47 & $15.8^{+5.9}_{-3.8}$ & $4.8^{+5.1}_{-2.0}$ & $140^{+56}_{-32}$ & $474^{+842}_{-297}$ &  q? \\ 
31  & 174804.2-244700 & 17:48:04.241  0.004 & -24:47:00.80  0.07 & 18.06 & $5.7^{+4.5}_{-2.2}$ & $14.7^{+6.5}_{-3.6}$ & $107^{+47}_{-23}$ & $308^{+681}_{-235}$ &  V? \\ 
32  & 174805.3-244631 & 17:48:05.364  0.007 & -24:46:31.43  0.10 & 15.09 & $6.7^{+4.7}_{-2.5}$ & $11.6^{+6.1}_{-3.2}$ & $113^{+48}_{-26}$ & $72^{+32}_{-20}$ &  \\
33  & 174804.7-244650 & 17:48:04.767  0.007 & -24:46:50.93  0.07 & 6.03 & $8.6^{+5.0}_{-2.8}$ & $8.7^{+5.7}_{-2.7}$ & $82^{+40}_{-19}$ & $348^{+722}_{-249}$ &  \\ 
34  & 174804.7-244604 & 17:48:04.706  0.007 & -24:46:04.82  0.07 & 40.45 & $4.8^{+4.4}_{-2.0}$ & $11.7^{+6.1}_{-3.2}$ & $87^{+44}_{-21}$ & $309^{+656}_{-229}$ &  \\ 
35  & 174805.0-244652 & 17:48:05.018  0.006 & -24:46:52.87  0.08 & 7.84 & $2.9^{+3.9}_{-1.6}$ & $12.6^{+6.2}_{-3.4}$ & $107^{+48}_{-27}$ & $44^{+29}_{-16}$ &  \\
36  & 174805.6-244642 & 17:48:05.692  0.005 & -24:46:42.67  0.12 & 10.99 & $8.7^{+5.0}_{-2.8}$ & $6.7^{+5.4}_{-2.4}$ & $74^{+34}_{-19}$ & $51^{+28}_{-16}$ &  \\ 
37  & 174804.6-244652 & 17:48:04.609  0.007 & -24:46:52.34  0.08 & 8.23 & $2.7^{+3.9}_{-1.4}$ & $12.6^{+6.2}_{-3.3}$ & $90^{+47}_{-21}$ & $24^{+25}_{-10}$ & V  \\ 
38  & 174805.3-244656 & 17:48:05.391  0.008 & -24:46:56.28  0.08 & 12.92 & $3.9^{+4.1}_{-1.9}$ & $7.7^{+5.6}_{-2.5}$ & $66^{+41}_{-19}$ & $18^{+16}_{-8}$ &  \\
39  & 174804.9-244642 & 17:48:04.920  0.004 & -24:46:42.84  0.10 & 2.34 & $2.9^{+3.9}_{-1.6}$ & $7.8^{+5.6}_{-2.5}$ & $78^{+48}_{-24}$ & $25^{+21}_{-10}$ &  \\ 
40  & 174804.6-244625 & 17:48:04.651  0.006 & -24:46:25.16  0.10 & 20.32 & $0.9^{+3.5}_{-0.7}$ & $9.7^{+5.8}_{-2.9}$ & $57^{+34}_{-17}$ & $3^{+8}_{-2}$ &  \\ 
41  & 174804.2-244624 & 17:48:04.240  0.005 & -24:46:24.28  0.07 & 22.78 & $5.0^{+4.3}_{-2.1}$ & $4.9^{+5.2}_{-1.9}$ & $69^{+51}_{-21}$ & $31^{+24}_{-12}$ & \\ 
42  & 174804.8-244628 & 17:48:04.885  0.010 & -24:46:28.44  0.11 & 16.74 & $0.8^{+3.5}_{-0.7}$ & $8.6^{+5.7}_{-2.7}$ & $50^{+34}_{-15}$ & $17^{+16}_{-8}$ & \\ 
43  & 174804.0-244647 & 17:48:04.014  0.007 & -24:46:47.38  0.12 & 12.35 & $1.8^{+3.7}_{-1.2}$ & $7.6^{+5.5}_{-2.6}$ & $44^{+26}_{-14}$ & $10^{+14}_{-5}$ &  \\ 
44  & 174804.4-244632 & 17:48:04.432  0.011 & -24:46:32.89  0.12 & 13.89 & $2.8^{+3.9}_{-1.6}$ & $5.8^{+5.2}_{-2.3}$ & $47^{+32}_{-16}$ & $9^{+10}_{-5}$ &  \\ 
45  & 174805.2-244639 & 17:48:05.265  0.009 & -24:46:39.87  0.11 & 7.21 & $6.7^{+4.7}_{-2.5}$ & $1.7^{+4.6}_{-1.1}$ & $44^{+30}_{-15}$ & $35^{+25}_{-14}$ &  \\ 
46  & 174807.4-244658 & 17:48:07.446  0.008 & -24:46:58.50  0.07 & 37.07 & $2.9^{+3.9}_{-1.6}$ & $4.9^{+5.1}_{-2.0}$ & $42^{+34}_{-14}$ & $10^{+10}_{-6}$ & \\ 
47  & 174804.2-244606 & 17:48:04.249  0.008 & -24:46:06.66  0.09 & 39.54 & $1.9^{+3.7}_{-1.3}$ & $5.8^{+5.3}_{-2.2}$ & $44^{+31}_{-15}$ & $29^{+25}_{-12}$ &  \\ 
48  & 174806.3-244637 & 17:48:06.326  0.014 & -24:46:37.59  0.16 & 20.77 & $0.0^{+3.2}_{-0.0}$ & $6.7^{+5.4}_{-2.3}$ & $39^{+32}_{-14}$ & $4^{+11}_{-4}$ &  \\ 
49  & 174806.8-244644 & 17:48:06.824  0.009 & -24:46:44.39  0.16 & 26.13 & $1.9^{+3.7}_{-1.3}$ & $2.9^{+4.8}_{-1.5}$ & $31^{+30}_{-14}$ & $6^{+9}_{-4}$ &  \\ 
50  & 174805.8-244646 & 17:48:05.877  0.010 & -24:46:46.52  0.14 & 13.29 & $4.8^{+4.4}_{-1.9}$ & $0.0^{+4.2}_{-0.0}$ & $17^{+20}_{-8}$ & $289^{+655}_{-228}$ & q? \\
\enddata
\tablecomments{
Names, positions, distance from center of Terzan 5, counts in two X-ray energy bands (energies given
in keV),
and estimated X-ray luminosities (in units of $10^{30}$ ergs s$^{-1}$)
of X-ray sources associated with Terzan 5.  The errors in parentheses
after the position represent 
the $1\sigma$ uncertainties in the relative positions of the sources,
derived from ACIS\_EXTRACT centroiding.  The counts in each band are the
numbers of photons within the source regions of Figure
1. Luminosities are computed from the corrected photon fluxes in
several narrow bands, see text. Notes indicate short-term variability 
(V = 99\% confidence, V? = 95\% confidence),
years-timescale variability between 2000 and 2003 (Y), and possible
identifications (L = transient LMXB EXO 1745-248, q = qLMXB, q? = qLMXB 
candidate, s = foreground star, MSP = radio millisecond pulsar). 
}
\end{deluxetable}
\clearpage


\clearpage

\begin{deluxetable}{lcccccccccccr}
\tabletypesize{\footnotesize}
\tablewidth{7.5truein}
\tablecaption{\textbf{Serendipitous Sources in the Terzan 5 Field}}
\tablehead{
\colhead{\textbf{Name}} & \colhead{RA}  & \colhead{Dec}   &
\colhead{Distance} & \multicolumn{2}{c}{Counts} & \colhead{Flux}  
& \colhead{Op?} & \colhead{B} & \colhead{R} & \colhead{I} \\ 
(CXOU J) & (HH:MM:SS) & (DD:MM:SS) & (\arcsec)  & (0.5-2 keV) & (2-6 keV) & (0.5-6)
& & & & \\
}
\startdata
174822.8-244600 & 17:48:22.813  0.011 & -24:46:00.66  0.28 & 247.91 & $9.3^{+5.1}_{-2.9}$ & $17.2^{+6.7}_{-3.9}$ & $22.7^{+7.9}_{-4.4}$ &     \\ 
174821.5-244507 & 17:48:21.532  0.020 & -24:45:07.59  0.34 & 246.58 & $2.2^{+3.8}_{-1.3}$ & $1.5^{+4.5}_{-0.9}$ & $1.7^{+3.2}_{-0.9}$ &     \\ 
174821.0-244622 & 17:48:21.083  0.017 & -24:46:22.17  0.39 & 221.51 & $4.5^{+4.3}_{-2.0}$ & $6.4^{+5.4}_{-2.3}$ & $7.2^{+4.1}_{-2.1}$ &     \\ 
174820.7-244240 & 17:48:20.775  0.029 & -24:42:40.98  0.33 & 326.07 & $16.3^{+5.9}_{-3.9}$ & $3.2^{+4.8}_{-1.4}$ & $8.0^{+4.8}_{-2.5}$ &     \\ 
174818.1-244758 & 17:48:18.168  0.010 & -24:47:58.75  0.13 & 195.01 & $1.7^{+3.7}_{-1.0}$ & $4.6^{+5.1}_{-1.9}$ & $3.4^{+3.1}_{-1.2}$ &  c* & - & 17.2 & 15.9  \\
174818.0-244255 & 17:48:18.038  0.027 & -24:42:55.58  0.41 & 291.31 & $1.7^{+3.7}_{-1.1}$ & $17.9^{+6.7}_{-4.1}$ & $11.2^{+4.6}_{-2.4}$ &    \\
174817.7-244046 & 17:48:17.759  0.045 & -24:40:46.72  0.58 & 398.16 & $13.6^{+5.7}_{-3.5}$ & $0.8^{+4.4}_{-0.7}$ & $4.9^{+3.5}_{-1.8}$ &     \\ 
174815.7-244801 & 17:48:15.727  0.010 & -24:48:01.61  0.19 & 166.00 & $2.7^{+3.9}_{-1.4}$ & $1.7^{+4.5}_{-1.0}$ & $3.1^{+3.5}_{-1.3}$ &     \\ 
174815.5-244424 & 17:48:15.574  0.026 & -24:44:24.89  0.31 & 201.97 & $7.2^{+4.7}_{-2.5}$ & $0.8^{+4.4}_{-0.7}$ & $3.0^{+2.5}_{-1.0}$ &     \\ 
174814.9-244728 & 17:48:14.994  0.012 & -24:47:28.97  0.42 & 144.19 & $4.6^{+4.3}_{-2.0}$ & $0.8^{+4.4}_{-0.7}$ & $2.7^{+2.6}_{-1.4}$ &     \\ 
174814.7-244802 & 17:48:14.748  0.008 & -24:48:02.13  0.17 & 154.55 & $2.7^{+3.9}_{-1.4}$ & $16.4^{+6.7}_{-3.9}$ & $10.1^{+4.2}_{-2.3}$ &    \\ 
174813.5-244832 & 17:48:13.537  0.017 & -24:48:32.95  0.28 & 159.46 & $2.7^{+3.9}_{-1.4}$ & $0.9^{+4.4}_{-0.8}$ & $1.8^{+2.4}_{-0.8}$ &     \\ 
174812.7-244811 & 17:48:12.700  0.004 & -24:48:11.22  0.08 & 136.63 & $45.7^{+8.6}_{-6.7}$ & $3.8^{+4.9}_{-1.8}$ & $22.5^{+5.3}_{-4.1}$ & c* & 18.4 & 15.1 & -    \\
174811.9-244550 & 17:48:11.981  0.011 & -24:45:50.21  0.17 & 110.94 & $5.7^{+4.5}_{-2.3}$ & $8.6^{+5.7}_{-2.7}$ & $7.8^{+4.2}_{-2.0}$ &     \\ 
174811.3-244516 & 17:48:11.306  0.013 & -24:45:16.87  0.20 & 124.08 & $3.8^{+4.1}_{-1.9}$ & $5.8^{+5.3}_{-2.2}$ & $5.4^{+3.6}_{-1.8}$ &   c & 18.5 & 16.0 & 15.7 \\
174811.2-244656 & 17:48:11.209  0.008 & -24:46:56.22  0.20 & 86.54 & $4.8^{+4.3}_{-2.0}$ & $5.8^{+5.2}_{-2.2}$ & $7.4^{+4.4}_{-2.3}$ &     \\
174810.9-244421 & 17:48:10.966  0.023 & -24:44:21.42  0.12 & 165.77 & $1.6^{+3.7}_{-1.0}$ & $0.9^{+4.4}_{-0.8}$ & $1.9^{+3.2}_{-1.1}$ &    \\ 
174810.4-244235 & 17:48:10.491  0.020 & -24:42:35.13  0.25 & 261.33 & $26.2^{+7.1}_{-5.0}$ & $5.9^{+5.3}_{-2.2}$ & $13.0^{+4.5}_{-2.9}$ &   \\ 
174810.3-244845 & 17:48:10.322  0.018 & -24:48:45.44  0.19 & 141.07 & $1.8^{+3.7}_{-1.1}$ & $1.8^{+4.6}_{-1.1}$ & $2.0^{+2.7}_{-0.9}$ &     \\ 
174810.1-244902 & 17:48:10.132  0.013 & -24:49:02.39  0.11 & 154.57 & $1.9^{+3.7}_{-1.1}$ & $5.9^{+5.3}_{-2.2}$ & $12.7^{+9.1}_{-4.4}$ &    \\ 
174809.7-244437 & 17:48:09.729  0.011 & -24:44:37.93  0.19 & 143.20 & $16.5^{+6.0}_{-3.9}$ & $2.7^{+4.7}_{-1.4}$ & $7.9^{+3.7}_{-2.3}$ &  c* & 19.4 & 17.7 & 16.7  \\ 
174809.6-244640 & 17:48:09.691  0.016 & -24:46:40.98  0.16 & 65.10 & $5.9^{+4.5}_{-2.3}$ & $0.9^{+4.4}_{-0.7}$ & $4.9^{+3.6}_{-1.9}$ &     \\ 
174808.8-244630 & 17:48:08.878  0.018 & -24:46:30.55  0.18 & 56.04 & $4.8^{+4.3}_{-2.0}$ & $0.9^{+4.4}_{-0.8}$ & $3.5^{+2.8}_{-1.4}$ &     \\ 
174808.7-244507 & 17:48:08.744  0.022 & -24:45:07.59  0.30 & 110.70 & $6.6^{+4.7}_{-2.4}$ & $0.0^{+4.2}_{-0.0}$ & $2.2^{+2.3}_{-1.1}$ &     \\ 
174808.7-244648 & 17:48:08.736  0.011 & -24:46:48.57  0.18 & 52.26 & $1.7^{+3.7}_{-1.1}$ & $3.8^{+4.9}_{-1.8}$ & $3.2^{+3.0}_{-1.2}$ &     \\ 
174808.6-244101 & 17:48:08.687  0.040 & -24:41:01.23  0.46 & 348.37 & $7.9^{+4.7}_{-2.7}$ & $7.2^{+5.5}_{-2.5}$ & $4.0^{+2.7}_{-1.4}$ &    \\ 
174808.1-244757 & 17:48:08.184  0.016 & -24:47:57.51  0.19 & 85.00 & $3.8^{+4.1}_{-1.8}$ & $3.7^{+4.9}_{-1.7}$ & $4.2^{+3.7}_{-1.4}$ &     \\ 
174807.2-244857 & 17:48:07.289  0.012 & -24:48:57.09  0.22 & 135.85 & $2.9^{+4.0}_{-1.5}$ & $4.8^{+5.1}_{-2.0}$ & $5.8^{+4.3}_{-1.9}$ &    \\ 
174807.0-244328 & 17:48:07.043  0.023 & -24:43:28.64  0.25 & 198.50 & $2.9^{+4.0}_{-1.4}$ & $8.2^{+5.6}_{-2.6}$ & $6.1^{+4.2}_{-1.9}$ &    \\ 
174806.4-244226 & 17:48:06.467  0.025 & -24:42:26.00  0.33 & 259.95 & $13.3^{+5.6}_{-3.5}$ & $1.2^{+4.5}_{-0.7}$ & $3.4^{+3.7}_{-1.4}$ &  ? & 16.4 & 14.0 & 13.5   \\
174806.3-244504 & 17:48:06.396  0.011 & -24:45:04.78  0.14 & 102.34 & $0.0^{+3.2}_{-0.0}$ & $12.8^{+6.2}_{-3.4}$ & $12.0^{+5.7}_{-3.3}$ &    \\
174806.1-244806 & 17:48:06.194  0.005 & -24:48:06.14  0.07 & 82.84 & $7.6^{+4.8}_{-2.6}$ & $16.7^{+6.7}_{-3.9}$ & $12.3^{+4.7}_{-2.5}$ &     \\ 
174805.8-244534 & 17:48:05.821  0.004 & -24:45:34.92  0.04 & 71.35 & $51.8^{+9.1}_{-7.1}$ & $10.8^{+5.9}_{-3.1}$ & $24.8^{+6.0}_{-4.2}$ &  c & 18.6 & 16.0 & 14.5    \\
174805.4-244333 & 17:48:05.461  0.013 & -24:43:33.30  0.27 & 192.03 & $4.5^{+4.3}_{-2.0}$ & $5.5^{+5.2}_{-2.1}$ & $5.5^{+3.4}_{-1.6}$ &     \\
174804.4-244503 & 17:48:04.496  0.006 & -24:45:03.19  0.07 & 102.14 & $9.8^{+5.1}_{-3.0}$ & $20.7^{+7.1}_{-4.4}$ & $18.5^{+6.0}_{-3.3}$ &   \\ 
174804.4-244543 & 17:48:04.459  0.005 & -24:45:43.19  0.08 & 62.28 & $4.9^{+4.4}_{-2.0}$ & $6.8^{+5.4}_{-2.4}$ & $6.5^{+3.6}_{-1.8}$ &     \\ 
174804.2-244302 & 17:48:04.228  0.028 & -24:43:02.32  0.28 & 223.05 & $3.2^{+4.0}_{-1.6}$ & $6.6^{+5.4}_{-2.3}$ & $5.8^{+4.1}_{-1.7}$ &    \\ 
174803.3-244854 & 17:48:03.370  0.004 & -24:48:54.23  0.05 & 130.74 & $77.6^{+10.6}_{-8.7}$ & $0.9^{+4.4}_{-0.8}$ & $20.7^{+5.1}_{-3.9}$ & c* & 15.3 & 13.2 & 13.2  \\
174803.3-244749 & 17:48:03.321  0.014 & -24:47:49.53  0.16 & 67.88 & $5.7^{+4.5}_{-2.2}$ & $2.8^{+4.8}_{-1.4}$ & $5.2^{+3.9}_{-1.7}$ &     \\
174802.3-244445 & 17:48:02.352  0.002 & -24:44:45.93  0.05 & 124.22 & $1.7^{+3.7}_{-1.0}$ & $0.0^{+4.2}_{-0.0}$ & $0.3^{+0.8}_{-0.3}$ &     \\ 
174801.6-244747 & 17:48:01.656  0.011 & -24:47:47.49  0.15 & 76.47 & $5.6^{+4.5}_{-2.2}$ & $8.6^{+5.7}_{-2.7}$ & $7.9^{+4.3}_{-2.1}$ &    \\ 
174801.5-244440 & 17:48:01.583  0.021 & -24:44:40.50  0.12 & 132.64 & $0.9^{+3.5}_{-0.8}$ & $4.7^{+5.1}_{-2.0}$ & $2.9^{+2.3}_{-1.2}$ &    \\ 
174801.5-244621 & 17:48:01.527  0.007 & -24:46:21.13  0.23 & 51.93 & $9.7^{+5.2}_{-2.9}$ & $3.6^{+4.9}_{-1.6}$ & $4.1^{+3.2}_{-1.4}$ &     \\ 
174801.3-244815 & 17:48:01.323  0.010 & -24:48:15.35  0.26 & 102.53 & $5.7^{+4.5}_{-2.2}$ & $0.0^{+4.2}_{-0.0}$ & $1.0^{+2.0}_{-0.7}$ &    \\ 
174759.9-244734 & 17:47:59.982  0.015 & -24:47:34.05  0.21 & 82.92 & $6.8^{+4.7}_{-2.5}$ & $2.8^{+4.8}_{-1.5}$ & $4.6^{+3.1}_{-1.7}$ &   c & 15.9 & 13.5 & 10.9  \\
174759.8-244529 & 17:47:59.810  0.008 & -24:45:29.56  0.24 & 102.84 & $4.7^{+4.3}_{-2.0}$ & $1.8^{+4.6}_{-1.1}$ & $4.0^{+3.1}_{-1.5}$ &  ? & 19.43 & - & 17.61 \\
174759.7-244504 & 17:47:59.753  0.039 & -24:45:04.32  0.24 & 122.88 & $2.8^{+3.9}_{-1.5}$ & $1.9^{+4.6}_{-1.1}$ & $3.8^{+4.1}_{-1.6}$ &     \\ 
174759.6-244811 & 17:47:59.641  0.015 & -24:48:11.89  0.13 & 112.52 & $0.0^{+3.2}_{-0.0}$ & $2.7^{+4.8}_{-1.3}$ & $2.0^{+3.0}_{-1.0}$ &     \\ 
174759.1-244610 & 17:47:59.113  0.011 & -24:46:10.43  0.11 & 86.22 & $7.7^{+4.9}_{-2.6}$ & $2.8^{+4.8}_{-1.5}$ & $5.3^{+3.2}_{-1.8}$ &   c* & 18.85 & 14.8 & 13.2 \\ 
174758.7-244228 & 17:47:58.783  0.022 & -24:42:28.23  0.24 & 270.07 & $4.7^{+4.3}_{-2.0}$ & $20.1^{+7.1}_{-4.3}$ & $14.5^{+5.2}_{-2.8}$ &     \\ 
174758.7-244429 & 17:47:58.743  0.013 & -24:44:29.11  0.32 & 159.88 & $1.6^{+3.7}_{-1.0}$ & $3.7^{+4.9}_{-1.8}$ & $2.9^{+2.8}_{-1.1}$ &     \\ 
174758.3-244837 & 17:47:58.353  0.015 & -24:48:37.59  0.17 & 143.53 & $3.8^{+4.2}_{-1.8}$ & $0.0^{+4.2}_{-0.0}$ & $0.8^{+1.9}_{-0.7}$ &     \\ 
174757.2-244122 & 17:47:57.228  0.035 & -24:41:22.29  0.32 & 338.94 & $4.6^{+4.3}_{-1.9}$ & $17.9^{+6.9}_{-4.0}$ & $16.5^{+6.2}_{-3.5}$ &    \\ 
174757.0-244615 & 17:47:57.041  0.024 & -24:46:15.49  0.17 & 111.15 & $2.8^{+3.9}_{-1.5}$ & $3.8^{+4.9}_{-1.7}$ & $4.3^{+3.7}_{-1.5}$ &     \\ 
174756.9-244218 & 17:47:56.988  0.032 & -24:42:18.44  0.25 & 287.56 & $1.4^{+3.6}_{-0.9}$ & $20.0^{+7.1}_{-4.3}$ & $15.5^{+5.7}_{-3.4}$ &   \\ 
174755.8-244622 & 17:47:55.855  0.010 & -24:46:22.69  0.35 & 125.31 & $0.9^{+3.5}_{-0.7}$ & $2.8^{+4.8}_{-1.5}$ & $1.8^{+2.1}_{-0.9}$ &     \\ 
174755.1-244154 & 17:47:55.120  0.023 & -24:41:54.51  0.23 & 319.71 & $35.3^{+7.8}_{-5.8}$ & $7.8^{+5.6}_{-2.5}$ & $26.1^{+6.5}_{-4.4}$ &   c & 19.4 & 14.7 & 13.5  \\
174754.2-244630 & 17:47:54.290  0.020 & -24:46:30.00  0.16 & 145.38 & $10.5^{+5.2}_{-3.1}$ & $0.0^{+4.2}_{-0.0}$ & $2.6^{+2.5}_{-1.3}$ &   c* & 12.6 & 11.7 & 10.3  \\ 
174754.2-244444 & 17:47:54.217  0.024 & -24:44:44.32  0.24 & 189.56 & $1.8^{+3.7}_{-1.2}$ & $6.5^{+5.3}_{-2.4}$ & $4.7^{+3.5}_{-1.5}$ &     \\
174754.0-244335 & 17:47:54.074  0.022 & -24:43:35.43  0.17 & 240.36 & $2.2^{+3.8}_{-1.2}$ & $11.7^{+6.1}_{-3.2}$ & $8.0^{+4.4}_{-2.1}$ &    \\ 
174752.6-244648 & 17:47:52.679  0.015 & -24:46:48.39  0.27 & 166.55 & $0.0^{+3.2}_{-0.0}$ & $5.5^{+5.2}_{-2.1}$ & $2.9^{+2.4}_{-1.1}$ &     \\ 
174752.5-244457 & 17:47:52.522  0.002 & -24:44:57.67  0.03 & 200.03 & $0.8^{+3.5}_{-0.7}$ & $0.8^{+4.3}_{-0.7}$ & $0.9^{+2.7}_{-0.8}$ &     \\ 
174752.5-244247 & 17:47:52.518  0.019 & -24:42:47.03  0.24 & 291.45 & $16.8^{+6.0}_{-4.0}$ & $16.3^{+6.6}_{-3.8}$ & $18.6^{+5.7}_{-3.2}$ &   \\ 
174751.7-244657 & 17:47:51.759  0.009 & -24:46:57.45  0.07 & 179.47 & $8.7^{+4.9}_{-2.8}$ & $22.5^{+7.3}_{-4.6}$ & $18.2^{+5.7}_{-3.3}$ &    \\ 
174751.7-244356 & 17:47:51.751  0.018 & -24:43:56.36  0.16 & 246.19 & $21.8^{+6.6}_{-4.5}$ & $7.4^{+5.5}_{-2.5}$ & $13.0^{+4.7}_{-2.9}$ &  c  & 17.4 & - & 14.5 \\ 
174751.2-244620 & 17:47:51.233  0.010 & -24:46:20.77  0.09 & 187.81 & $3.7^{+4.1}_{-1.8}$ & $0.8^{+4.4}_{-0.7}$ & $2.2^{+2.8}_{-1.0}$ &     \\ 
174750.4-244615 & 17:47:50.431  0.018 & -24:46:15.66  0.12 & 199.34 & $13.2^{+5.6}_{-3.5}$ & $3.5^{+4.9}_{-1.6}$ & $8.4^{+4.5}_{-2.4}$ &   c & 19.7 & 16.7 & 13.9  \\ 
174750.3-244638 & 17:47:50.377  0.031 & -24:46:38.83  0.31 & 197.76 & $6.3^{+4.6}_{-2.3}$ & $2.6^{+4.7}_{-1.4}$ & $4.2^{+3.9}_{-1.8}$ &     \\
174749.9-244445 & 17:47:49.944  0.033 & -24:44:45.54  0.34 & 235.68 & $5.6^{+4.5}_{-2.2}$ & $1.7^{+4.6}_{-1.1}$ & $2.1^{+3.1}_{-1.0}$ &     \\ 
174749.8-244628 & 17:47:49.831  0.025 & -24:46:28.05  0.18 & 206.02 & $6.4^{+4.6}_{-2.4}$ & $2.6^{+4.7}_{-1.4}$ & $5.2^{+3.9}_{-1.7}$ &     \\ 
174749.7-244351 & 17:47:49.754  0.026 & -24:43:51.57  0.27 & 269.55 & $6.4^{+4.6}_{-2.3}$ & $2.9^{+4.8}_{-1.4}$ & $6.5^{+4.4}_{-2.1}$ &  ? & 19.4 & 19.7 & 17.7   \\
174749.6-244637 & 17:47:49.687  0.038 & -24:46:37.90  0.32 & 207.34 & $1.7^{+3.7}_{-1.0}$ & $6.6^{+5.4}_{-2.4}$ & $6.0^{+4.3}_{-2.1}$ &     \\
174749.5-244719 & 17:47:49.567  0.006 & -24:47:19.51  0.55 & 211.70 & $0.0^{+3.2}_{-0.0}$ & $2.5^{+4.7}_{-1.4}$ & $1.4^{+1.9}_{-0.8}$ &     \\ 
174749.2-244619 & 17:47:49.293  0.013 & -24:46:19.40  0.09 & 214.19 & $11.0^{+5.3}_{-3.2}$ & $31.2^{+8.1}_{-5.4}$ & $24.6^{+6.1}_{-3.8}$ &   \\ 
174748.6-244757 & 17:47:48.622  0.030 & -24:47:57.70  0.50 & 233.31 & $0.7^{+3.4}_{-0.7}$ & $2.6^{+4.7}_{-1.3}$ & $2.6^{+3.8}_{-1.2}$ &     \\ 
174747.4-244806 & 17:47:47.408  0.047 & -24:48:06.09  0.27 & 251.65 & $3.4^{+4.1}_{-1.7}$ & $2.7^{+4.7}_{-1.4}$ & $4.9^{+4.2}_{-1.9}$ &    \\ 
174746.2-244519 & 17:47:46.234  0.012 & -24:45:19.80  0.13 & 268.79 & $6.6^{+4.6}_{-2.4}$ & $39.4^{+8.7}_{-6.1}$ & $69.2^{+15.4}_{-10.7}$ &   \\ 
\enddata		    			     		    
\tablecomments{Sources outside the Terzan 5 half-mass radius detected
  on the S3 chip.  Relative positional errors are given in parentheses on
the last quoted digits.  Energy flux in units of $10^{15}$ ergs 
cm$^{-2}$ s$^{-1}$, assuming $N_H=1.2\times10^{22}$ cm$^{-2}$.
 Notes (Op?) indicate likely optical counterparts
(``c''), and less likely optical counterparts (``?''); those used to align 
the X-ray positions are indicated (``*''). Magnitudes from 
USNO B1.0 digitized sky survey \citep{Monet03}, averaged when two 
measurements exist; note that there is substantial scatter among these 
measurements (up to 1 mag), presumably due to crowding.
}
\end{deluxetable}

\clearpage

\begin{landscape}

\begin{deluxetable}{lccccccccccccr}
\tabletypesize{\footnotesize}
\tablewidth{8.5truein}
\tablecaption{\textbf{Spectral Fits to Brighter Terzan 5 Sources}}
\tablehead{
\colhead{\textbf{Source}} & \multicolumn{5}{c}{H-atmosphere + Power-law}  &
\multicolumn{4}{c}{MEKAL} & 
 \multicolumn{4}{c}{Power-law}  \\
\cline{2-6} \cline{11-14} \\
 & (kT, eV) & ($N_H\times10^{22}$) & (PL fraction) &
 ($\chi^2_{\nu}$/dof) & ($L_X$,$10^{30}$ ) 
& (kT, keV) & ($N_H\times10^{22}$) & ($\chi^2_{\nu}$/dof) &
 ($L_X$,$10^{30}$ )  
& ($\alpha$) & ($N_H\times10^{22}$) & ($\chi^2_{\nu}$/dof)  & ($L_X$,
 $10^{30}$)  
}
\startdata

CX1 & 
83$^{+56}_{-83}$ & 1.91$^{+0.87}_{-0.32}$ & 94$^{+6}_{-34}$ & 0.98/36 & 2.9e33 
&  $>11.5$       & 1.52$^{+0.48}_{-0.26}$ & 0.93/36  & 2.3e33
& 1.10$^{+0.39}_{-0.26}$ & 1.33$^{+0.66}_{-0.13}$ & 0.92/36 & 2.1e33 \\
CX2 &  
143$^{+8}_{-8}$ & 1.49$^{+0.23}_{-0.16}$ & 28$^{+8}_{-7}$  & 1.21/28 & 3.2e33 
& 2.5$^{+0.5}_{-0.4}$  &  1.2$^{+0.09}_{-0}$  &  1.84/28  &   1.8e33
& 2.70$^{+0.57}_{-0.27}$  & 1.23$^{+0.38}_{-0.01}$  & 1.43/28 & 2.7e33  \\
CX3 &
107$^{+24}_{-23}$ & 1.28$^{+0.55}_{-0.08}$ & 68$^{+22}_{-24}$ & 0.72/23 & 1.9e33 
& 5.3$^{+3.6}_{-1.6}$ & 1.20$^{+0.22}_{-0}$ & 0.75/23 & 1.4e33 
& 1.86$^{+0.33}_{-0.25}$ & 1.20$^{+0.33}_{-0}$ & 0.72/23 & 1.6e33 \\
CX4 &
81$^{+53}_{-81}$ & 1.20$^{+0.67}_{-0}$ & 89$^{+11}_{-48}$ & 0.92/19 & 1.32e33
& 10$^{+32}_{-5}$ & 1.20$^{+0.37}_{-0}$ & 0.97/19 & 1.16e33
& 1.59$^{+0.41}_{-0.25}$ & 1.20$^{+0.51}_{-0}$ & 0.93/19 & 1.23e33 \\
CX5 & 
62$^{+51}_{-62}$ & 1.37$^{+1.02}_{-0.17}$ & 97$^{+3}_{-34}$ & 0.67/18 & 1.31e33
& 6.6$^{+73}_{-3.4}$ & 1.58$^{+0.85}_{-0.38}$ & 0.65/18 & 1.37e33 
& 1.63$^{+0.67}_{-0.43}$ & 1.51$^{+0.96}_{-0.31}$ & 0.67/18 & 1.39e33 \\
CX6 & 
103$^{+25}_{-22}$ & 1.78$^{+0.80}_{-0.58}$ & 63$^{+22}_{-25}$ & 0.75/10 & 1.35e33
& 20$^{+60}_{-6}$ & 1.20$^{+0.39}_{-0.0}$ & 0.77/10 & 7.7e32
& 1.42$^{+0.49}_{-0.26}$ & 1.20$^{+0.60}_{-0.0}$ & 0.76/10 & 7.9e32 \\
CX7 & 
 9$^{+129}_{-9}$ & 2.68$^{+0.85}_{-0.62}$ & 100$^{+0}_{-67}$ & 0.97/8 & 9.9e32
& 9.1$^{+71}_{-5.8}$ & 2.84$^{+2.15}_{-1.06}$ & 0.96/8 & 1.0e33
& 1.87$^{+1.25}_{-0.48}$ & 3.29$^{+2.5}_{-1.7}$ & 0.93/8 & 1.34e33 \\
CX8 & 
116$^{+32}_{-116}$ & 2.92$^{+1.57}_{-1.33}$ & 53$^{+47}_{-27}$ & 1.01/8 & 1.85e33
& $>5.4$ & 1.75$^{+1.05}_{-0.48}$ & 0.96/8 & 7.9e32
& 0.91$^{+0.63}_{-0.45}$ & 1.36$^{+1.22}_{-0.16}$ & 0.90/8 & 6.8e32 \\
CX9 &
130$^{+13}_{-7}$ & 1.97$^{+0.50}_{-0.27}$ & 20$^{+5}_{-6}$ & 0.72/6 & 1.85e33
& 1.26$^{+1.44}_{-0.30}$ & 1.88$^{+0.67}_{-0.68}$ & 0.95/6 & 1.43e33
& 3.62$^{+1.88}_{-1.09}$ & 1.92$^{+1.28}_{-0.72}$ & 0.75/6 & 2.8e33 \\
CX10 &
72$^{+27}_{-72}$ & 1.90$^{+3.5}_{-0.63}$ & 89$^{+11}_{-28}$ & 0.86/6 & 7.5e32
 & $>3.1$ & 1.54$^{+1.59}_{-0.34}$ & 0.84/6 & 5.8e32 
& 1.25$^{+1.23}_{-0.59}$ & 1.49$^{+1.89}_{-0.29}$ & 0.84/6 & 5.7e32 \\
CX11$^a$ & 
75 & 2.12 & 85 & 3.32/5 & 7.1e32
& 80 & 1.64 & 3.25/5 & 5.1e32
& 0.78 & 1.2 & 3.06/5 & 4.5e32 \\
CX12 &
117$^{+11}_{-12}$ & 1.42$^{+0.46}_{-0.22}$ & 17$^{+16}_{-11}$ & 0.74/5 & 1.1e33
& 1.61$^{+0.52}_{-0.43}$ & 1.20$^{+0.45}_{-0}$ & 1.16/5 & 5.9e32 
& 3.26$^{+1.27}_{-0.48}$ & 1.20$^{+0.79}_{-0}$ & 0.70/5 & 1.1e33 \\
CX13 & 
17$^{+143}_{-17}$ & 3.80$^{+1.45}_{-1.00}$ & 100$^{+0}_{-83}$ & 0.66/5 & 7.7e32 
& 3.3$^{+\infty}_{-2.0}$ & 5.42$^{+4.8}_{-2.6}$ & 0.37/5 & 1.3e33
& 3.02$^{+2.38}_{-1.57}$ & 6.80$^{+6.06}_{-3.71}$ & 0.29/5 & 4.4e33 \\
CX14 &
128$^{+17}_{-19}$ & 2.47$^{+0.75}_{-0.48}$ & 20$^{+22}_{-13}$ & 0.62/5 & 1.8e33
& 3.9$^{+6.6}_{-2.1}$ & 1.20$^{+0.79}_{-0.0}$ & 1.19/5 & 4.6e32
& 2.28$^{+0.85}_{-0.72}$ & 1.41$^{+1.42}_{-0.21}$ & 1.11/5 & 6.4e32 \\
CX15 &
109$^{+12}_{-9}$ & 1.28$^{+0.46}_{-0.08}$ & 19$^{+15}_{-13}$ & 0.29/5 & 8.1e32
& 1.27$^{+0.55}_{-0.53}$ & 1.28$^{+0.62}_{-0.08}$ & 1.01/5 & 5.7e32
& 3.45$^{+1.47}_{-0.64}$ & 1.20$^{+0.76}_{-0}$ & 0.70/5 & 9.1e32 \\
CX16 & 
9$^{+129}_{-9}$ & 1.21$^{+0.66}_{-0.01}$ & 100$^{+0}_{-84}$ & 0.77/4 & 3.7e32
& 3.9$^{+\infty}_{-2.5}$ & 1.66$^{+1.63}_{-0.46}$ & 0.62/4 & 4.5e32 
& 2.17$^{+2.15}_{-1.11}$ & 1.82$^{+2.58}_{-0.62}$ & 0.66/4 & 5.8e32 \\
\enddata
\tablecomments{Spectral fits to cluster sources, with background
subtraction, in XSPEC.  Errors are 90\% confidence for a single
parameter; spectra are binned 
with 10 counts/bin for all sources 
with more than 70 counts, 8 counts/bin for sources with fewer
counts.  
All fits include photoelectric absorption
forced to be $\geq1.2\times10^{22}$ cm$^{-2}$, the cluster $N_H$
derived from near-infrared studies \citep{Cohn02}, plus dust
scattering for an assumed $A_V=6.7$.   For hydrogen
atmosphere plus power-law fits, we fixed the power-law photon index to
1.5, the NS radius to 10 km, and the NS mass to 1.4 \Msun. 
$^a$ CX11's spectrum does not produce acceptable fits with any of these models.
}
\end{deluxetable}

\clearpage

\end{landscape}


\end{document}